\documentclass{aa}
\usepackage{natbib}
\bibpunct{(}{)}{;}{a}{}{,} 
\usepackage{setspace}
\usepackage{amsmath}
\usepackage{graphicx}
\usepackage{tabularx}
\usepackage{lscape}
\usepackage{epstopdf}
\usepackage{longtable}
\usepackage{mathabx}

\begin{document}
\title{Combining direct imaging and radial velocity data towards a full exploration of the giant planet population}
\subtitle{I. Method and first results}

 

\author{
  J. Lannier \inst{1}
\and  A.M. Lagrange \inst{1}
\and  M. Bonavita \inst{2}
\and   S. Borgniet \inst{1}
\and   P. Delorme \inst{1}
\and N. Meunier \inst{1}
\and S. Desidera \inst{3}
\and S. Messina \inst{4}
\and G. Chauvin \inst{1}
\and M. Keppler \inst{1}
    }
\offprints{J. Lannier,
  \email{justine.lannier@univ-grenoble-alpes.fr}
   }
\institute{Univ. Grenoble Alpes, Institut de Plan\'etologie et d'Astrophysique de Grenoble (IPAG, UMR 5274), F-38000 Grenoble, France
\and Institute for Astronomy, The University of Edinburgh, Royal Observatory, Blackford Hill, Edinburgh, EH9 3HJ, U.K.
\and INAF Osservatorio Astornomico di Padova, vicolo dell’Osservatorio 5, I-35122, Padova, Italy
\and INAF- Catania Astrophysical Observatory, via S.Sofia, 78 I-95123 Catania, Italy}

\abstract{Thanks to the detections of more than 3000 exoplanets these last 20 years, statistical studies have already highlighted some properties in the distribution of the planet parameters. Nevertheless, few studies have yet investigated the planet populations from short to large separations around the same star since this requires the use of different detection techniques that usually target different types of stars.}
{We wish to develop a tool that combines direct and indirect methods so as to correctly investigate the giant planet populations at all separations.}
{We developed the MESS2 code, a Monte Carlo simulation code combining radial velocity and direct imaging data obtained at different epochs for a given star to estimate the detection probability of giant planets spanning a wide range of physical separations. It is based on the generation of synthetic planet populations.}
{We apply MESS2 on a young M1-type, the nearby star AU~Mic observed with HARPS and NACO/ESO. We show that giant planet detection limits are significantly improved at intermediate separations ($\approx$20~au in the case of AU~Mic). We show that the traditional approach of analysing independently the RV and DI detection limits systematically overestimates the planet detection limits and hence planet occurrence rates. The use of MESS2 allows to obtain correct planet occurrence rates in statistical studies, making use of multi-epoch DI data and/or RV measurements. We also show that MESS2 can optimise the schedule of future DI observations.}
{}

\date{}


\maketitle

\section{Introduction}

More than 3000 extra-solar planets have been detected to date, mostly with radial velocity (RV) and transit techniques (http://www.exoplanet.eu). Beside these prolific indirect detection methods, about 30 young, wide-orbit, giant exoplanets or brown dwarfs have been discovered using direct imaging techniques (DI). The analyses of the DI or RV surveys have brought in particular new constraints on exoplanet formation and highlight correlations between giant planet formation with  stellar mass or stellar metallicity \citep[][]{Biller.2007, Lafreniere.2007, Johnson.2008, Chauvin.2010, Johnson.2010, Bonfils.2011, Delorme.2012a, Rameau.2013, Vigan.2012, Bowler.2015}. Most of the analyses rely on robust statistical tools based on Monte Carlo simulations or on Bayesian methods and fake planet injection, such as the MESS code described by \citet{Bonavita.2011}.
\\
The imaged exoplanets generally populate age, mass and semi-major axis domains (<1~Gyr, >2~M$_{Jup}$, >5~au) distinct from those of RV planets usually searched (around mature stars, all masses, <5-10~au)\footnote{Transiting exoplanets are found on average on shorter separations.}. As a result, it has not been possible so far to explore the whole range of possible planetary orbits around a given star. It has been showed, though, that young stars can also be probed by RV \citep[eg,][]{Lagrange.2013} to search for giant planets, and that, combined with DI studies, can allow a full exploration of the parameter space, as far as giant planets are concerned. This brings new observational constraints at different separations, especially because the improving performances of adaptive optics allow now to probe regions closer to the star than previously \citep[with notably SPHERE, GPI, HiCIAO, ][]{Beuzit.2006,Macintosh.2014,Suzuki.2010}, therefore allowing overlap between the different datasets. This combination of data should significantly improve the robustness of the statistical studies of planetary populations. However, the simple consideration of the best detection limits derived from RV and DI data is not optimal for this analysis. If we consider planets on eccentric orbits in the overlapping separation range, the RV method will preferably detect the planets near their periastron since the RV variations are important there, while DI will detect the planets near their apastron since their projected separation is the largest. So, the actual chance to detect such planets is closer in this instance to the sum of the detection probability computed for each technique, rather than the best detection probability computed either for RV or for DI. A tool that self-consistently estimates the detection limits of a combined dataset is therefore necessary to derive the optimal detection limits for the growing number of stars currently investigated by RV, DI and in the future, astrometry with GAIA \citep{Sozzetti.2014,Perryman.2014}.
\\
\\
We present in this article the MESS2 code, an advanced version of MESS that combines RV and DI data obtained at different epochs for individual stars. MESS2 can be used 1) to investigate the giant planet population at all separations in order to test the models of planet population synthesis, 2) to help improving the observational strategy.
\\
We detail the methods for the DI and RV data combination in Section~\ref{methods}. In Section~\ref{applications}, we apply the MESS2 code to AU~Mic, a young and nearby M1 dwarf, and we present a quantitative analysis of systematic biases of using independent analyses instead of our self-consistent tool. We present our conclusions and the associated new perspectives of our results in Section~\ref{conclusion}.

\section{MESS2 (Multi-epochs multi-purposes exoplanet simulation system): methods}
\label{methods}
MESS2 allows to derive statistical information on the presence of planets around individual stars, using both multi-epoch DI and RV data. It is basically an upgrade of the MESS code \citep{Bonavita.2011}, which is optimised for a single dataset. The MESS2 principles are described in the following sections.

\subsection{Detection probability with direct imaging data at different epochs.}
\label{DI}
The MESS2 code uses a Monte Carlo approach, similarly to MESS. Step 1 is common to MESS and MESS2, while steps 2 and 3 are specific to MESS2.
\begin{enumerate}
\item {\bf Definition of the parameters.} We generate a synthetic population of planets using distributions predicted by theoretical models \citep[][]{Mordasini.2009} or semi-analytical approaches as the one by \cite{Cumming.2008}. 

The code builds a grid of mass and semi-major axis (a), with a tunable step.  A same set of $N_{gen}$ orbital parameters (eccentricity, inclination, argument of periastron, time of periastron passage) is then generated for each point of the grid. For each of the $N_{gen}$ sets the projected positions of the planets ($x,y$) on the plane perpendicular to the line of sight are calculated, using the equations (1) to (7) \citep{Bonavita.2011}.

\begin{equation}
     \begin{aligned}
        & x=AX+FY \\
        & y=BX+GY 
     \end{aligned}
\end{equation}

\begin{equation}
     \begin{aligned}
        & X=\cos E-e \\
        & Y=\sqrt{1-e^2}\sin E
     \end{aligned}
\end{equation}

\begin{equation}
     \begin{aligned}
    & \rho=\sqrt{x^2+y^2} 
     \end{aligned}
\end{equation}

\begin{equation}
     \begin{aligned}
& A=a(\cos\omega\cos\Omega-\sin\omega\sin\Omega\cos i) \\
& B=a(\cos\omega\sin\Omega+\sin\omega\cos\Omega\cos i) \\
& F=a(-\sin\omega\cos\Omega-\cos\omega\sin\Omega\cos i) \\
& G=a(-\sin\omega\sin\Omega+\cos\omega\cos\Omega\cos i)
     \end{aligned}
\end{equation}
\\
\\
\begin{equation}
     \begin{aligned}
    & M=(\frac{t_{obs}-T_0}{{\cal P}})2\pi \\
    & E_0=M+e\sin M +\frac{e^2}{2}\sin 2M \\ 
    & M_0=E_0-e\sin E_0 \\ 
     \end{aligned}
\end{equation}

\begin{equation}
     \begin{aligned}
    & E=E_0+\frac{M-M_0}{1-e\cos E_0}
     \end{aligned}
\end{equation}

\begin{equation}
     \begin{aligned}
    & \tan\frac{\nu}{2}=\sqrt{\frac{1+e}{1-e}}\tan\frac{E}{2}
     \end{aligned}
\end{equation}
\\
\\

\noindent $X$ and $Y$ are the orbital coordinates, computed using the Thiele-Innes elements $A,B,F,G$ described in Eq.~(4). $\rho$ is the projected separation. $E$ is the eccentric anomaly, $e$ the eccentricity, $\omega$ the argument of periastron, $\Omega$ the longitude of the ascending node, $i$ the inclination, $M$ the mean anomaly, $T_0$ the time of periastron passage, $\nu$ the true anomaly, ${\cal P}$ the orbital period of the planet, and $t_{obs}$ the time of the observations. For each point of the [mass,a] grid, we calculate the position of each synthetic planet for all the available epochs of DI observation.

\item {\bf Definition of the detectability.} For every observational epoch, the DI detectability of each planet is tested by comparing its mass to the detection capability of the considered instrument. Here, we use detection limit maps \citep[for further detail see][]{Delorme.2012a,Lannier_stat.2016}. If a planet is detectable at least at one epoch, it is considered as detectable by DI.

\item {\bf Probability derivation.} The probability to detect a planet of a given mass with a given semi-major axis is derived from the number of detectable planets over the $N_{gen}$ generated ones. Then, we build a probability map.
\end{enumerate}

\subsection{Deriving detection probabilities with RV data solely}
\label{RV}
MESS2 uses a root mean square-based method (RMS) as well as a local power analysis \citep[LPA,][]{Meunier.2012}. The RMS method is based on the comparison of the dispersion of synthetic planets RVs with the dispersion of the observed RVs. The LPA method is based on the generation of periodograms of synthetic planet RV time series, that are compared with the periodogram of the observed RV data {within given orbital periods.}

\subsubsection{RMS-based method}
\begin{enumerate}
\item {\bf Definition of the parameters.} The same set of planets used for the DI data analysis is used. We restrict however the semi-major axis range to values that correspond to periods less than half the available time baseline.
\item {\bf Definition of the detectability.} MESS2 computes synthetic RV curves for each synthetic planet. For a specific planet, given its mass $m_{\rm planet}$ (in $M_{Jup}$), its orbital period ${\cal P}$ (in year), its eccentricity $e$, and its inclination $i$, MESS2 computes the half amplitude $K$ (in m.s$^{-1}$) of the star RV variations (given the stellar mass $M_{\rm star}$, in $M_{\rm Sun}$), as shown in Eq.~\ref{eqEight}:

\begin{equation}
\label{eqEight}
K=\frac{28.432 \times m_{\rm planet} \times sin(i)}{M_{\rm star}^{2/3} \times {\cal P}^{1/3} \times \sqrt{1-e^2}}
\end{equation}
Then, the RV time series $V$ of the star due to the presence of the simulated planet orbiting the star are computed at each epoch when RV data were available, using:
\begin{equation}
V(t)=V_0(t)+K(t)\times(cos(\omega _*(t) +v_*(t))+e\times cos(\omega _*(t)))        
\end{equation}
$V_0$ is a white noise derived from the standard deviation of the observed temporal series of the star. $\omega _*$ is the argument of the periastron of the orbit of the star, $v_*$ the true anomaly of the star's position from the centre of mass, and $e$ the planet eccentricity. $N_{gen}$ such temporal series are generated for each point of the [mass,a] grid.
\\
To test the detectability of the generated planets, we compare the generated time series with the observed one and we apply the condition that no more than 0.15\% of the planet detections are actually false positive signals (this condition corresponds to a 3-sigma criterion). In practice, to fulfil the 3-sigma criterion, we generate a large number $N$ of RV time series without planet signal, by adding a white noise on the observed RV measurements, and we count how many times the RMS of the planet-less signal is above $X$ times the RMS of the observed RV, and we choose $X$ so that the number of false alarm detections does not exceed 0.15\% of $N$. We therefore consider that a planet is detected if the RMS of its simulated RV time series is higher than the observed one, using the threshold $X$:
\begin{equation}
\textrm{RMS}_{\textrm{generated RV}}>$X$\times \textrm{RMS}_{\textrm{observed RV}}
\end{equation}

\item {\bf Probability derivation.} The probability to detect planets of a given mass at a given semi-major axis is computed after counting how many planets are detected among the $N_{gen}$ generated ones, for each considered planetary mass and semi-major axis.
\end{enumerate}

\subsubsection{LPA-based method}
\begin{enumerate}
\item {\bf Definition of the parameters.} The free parameters for the generation of the periodograms are: i) the period range and the total number of periods for the computation of each periodogram; and ii) the orbital and physical parameters of the planets as described in Section~\ref{DI}. The width of the period windows used to compare the periodograms locally to test the detectability of the planets of period $P$ is fixed to [0.75$\times{\cal P}$;1.25$\times{\cal P}$] \citep[the choice of the width of the period window is discussed in][]{Meunier.2012}.
\item {\bf Definition of the detectability.} The Lomb-Scargle periodograms \citep{Lomb.1976,Scargle.1982,Press.1989} for each generated planet with a given mass and semi-major axis (hereafter called "synthetic periodograms") are computed, as well as the periodogram using the observed data. To determine the detectability of a planet, the maximum power in a synthetic periodogram is compared with the maximum power in the periodogram of the observed RV data, within the considered period window [0.75$\times{\cal P}$;1.25$\times{\cal P}$]. The periods {\cal P} are calculated from the input semi-major axes using the approximate third Kepler law\footnote{Note that since the conversion semi-major axis to period is made using the approximate third Kepler's law, the planet mass is neglected in this calculation. To mitigate the effects of this approximation when the mass ratio between the star and the planet is high, the value of the exact period (using the non-approximated formula of the third Kepler's law) is compared to the values of the approximated periods. The closest reference periodogram in terms of period is therefore used to derive the $N_{gen}$ periodograms for each point of the [mass,a] grid. We estimate that this period allocation leads to semi-major axis shifts of less than 4\%.}. 
 A planet is detected if its maximum power is at least 1.3 times higher than the maximum power in the periodogram of the observed RV data. The choice of this threshold results from an empirical compromise between our need to detect as many small planets as possible and our need to reduce as much as possible the rate of false positives. 
 Note that the rate of false positives is very difficult to estimate: as discussed in \citet{Meunier.2012}, the use of false alarm probability, which is sometimes used, has a limited interest, as it cannot take into account the temporal structure of the noise\footnote{In order to compare our threshold with the FAP formalism, we performed tests on a series of RV made of white noise and with the same calendars as the actual observations. We computed a "local FAP" at each period range, the level corresponding to the 0.1\%, 1\% etc "detectability" level. We see that the threshold is equivalent to a local FAP in the range 0.1-1\%, depending on the considered periods. In the case of non white noise (e.g. with peaks of stellar origine), the local FAP could be below the peaks in the periodogram that are signal of stellar origine. The 1.3 threshold will be higher than the local FAP in this range, which is more conservative and more realistic.}.
\\
To speed up the computation, the periodograms are computed only for a reference planetary mass $m_{\rm ref}$ for each semi-major axis and for each input orbital parameter (we call it the reference periodogram, and we arbitrarily choose $m_{\rm ref}=1$~M$_{Jup}$). We therefore get $N_{gen}$ reference periodograms for each value of semi-major axis. The periodograms corresponding to the other masses $m_{\rm p}$, but to the same orbital parameters, are scaled from the reference periodograms following the formula that we demonstrate in appendix~A:
\begin{equation}
\tiny
{\cal P}(P)={\cal P}_{{\rm ref}}(P)\times(\frac{m_{\rm p}}{m_{\rm ref}}\times (\frac{m_{\rm star}+m_{\rm ref}}{m_{\rm star}+m_{\rm p}})^{2/3})^{2}
\end{equation}
where ${\cal P}$ are the periodograms corresponding to the other masses $m_{\rm p}$, and ${\cal P}_{{\rm ref}}$ are the reference periodograms computed for 1~M$_{jup}$.
\\
For both the LPA-based and RMS-based method, only planets with periods lower than two times the time baseline are considered.

\item {\bf Probability derivation.} A probability map is computed by adding the number of detected planets among the $N_{gen}$ generated ones, for each planetary mass and semi-major axis of the grid.
\end{enumerate}

\subsubsection{Comparison of the two RV methods used in MESS2}
\citet{Meunier.2012} compared different methods to determine detection limits from RV data. They conclude that the LPA-based method is very robust and "provides the most significant improvement on the RMS method". Indeed, conversely to the RMS method, LPA method takes into account the temporal structures of the stellar noise which is in most cases the dominant source of noise. The RMS-based method can however be used to get a first estimate of the detection probability as it requires a much smaller computational time. This is especially useful for cases where large sets of RV measurements are available.
\\
\\
The RV time series are often dominated by the effects of the stellar activity and/or pulsation. Correction for these effects should be applied before using MESS2. In the case of magnetic activity it is possible to perform corrections using the bisector or other activity criteria (as in the case of AU~Mic, see below). In the case of pulsations, the RV can be averaged over appropriate timing (see an illustration in Lagrange et al. 2016, in prep., in the case of the pulsating A-type star $\beta$~Pictoris).

\subsection{Combining RV measurements and DI data obtained at different epochs}
\label{combination_RV+DI}
For each point of the [mass,a] grid, the detectability of each generated planet is determined by checking if each planet can be detectable at least in one DI epoch (see Section~\ref{DI}) or in the RV ones (see Section~\ref{RV}). The MESS2 code derives the final probability map, by counting for each [mass,a], how many of the $N_{gen}$ generated planets are detected with either technique.
\\
\\
Note: We set the detectability criterion at $X=1.44$ for the RMS-based method in the case of AU~Mic. This criteria corresponds to an equivalent 3-sigma which is appropriate for RV detections. For DI, we adopt the usual 5-sigma criterion: this criterion is realistic far from the star halo (in background limited region) but is probably optimistic in the speckle dominated region \citep[eg ][]{Delorme.2012a,Rameau.2013,Lannier_stat.2016}.

\section{Applications: the case of AU~Mic}
\label{applications}
AU~Mic (HIP~102409/HD~19748/GJ~803) is a M1V type star, that belongs to the $\beta$~Pictoris moving group \citep[20-26~Myr, in this paper we will use 21~Myr,][]{Binks.2014}, located at 10~pc \citep{Vanleeuwen.2007}. Its magnitude is $K=4.5$ and its mass is estimated to be 0.61~$M_{Sun}$ using BT-Settl models \citep{Allard.2012}. A debris disk was detected around AU~Mic from $\approx$10-17 to 210~au \citep{Liu.2004,Metchev.2005,Krist.2005}, with an edge-on configuration whose the inner part is asymmetric \citep{Kalas.2004,Liu.2004_nat,Boccaletti.2015}. The gas to dust mass ratio of the disk is low, 6:1 \citep{Roberge.2005}, compared to the typical mass ratio 100:1 of protoplanetary disks, indicating that most of the gas has dissipated. Neither planets nor brown dwarfs have been detected in the inner hole of the disk or within it yet.

\subsection{Observations}
\begin{figure*}[!h]
  \centering
   \begin{tabular}{cc}
    \includegraphics[width=90mm]{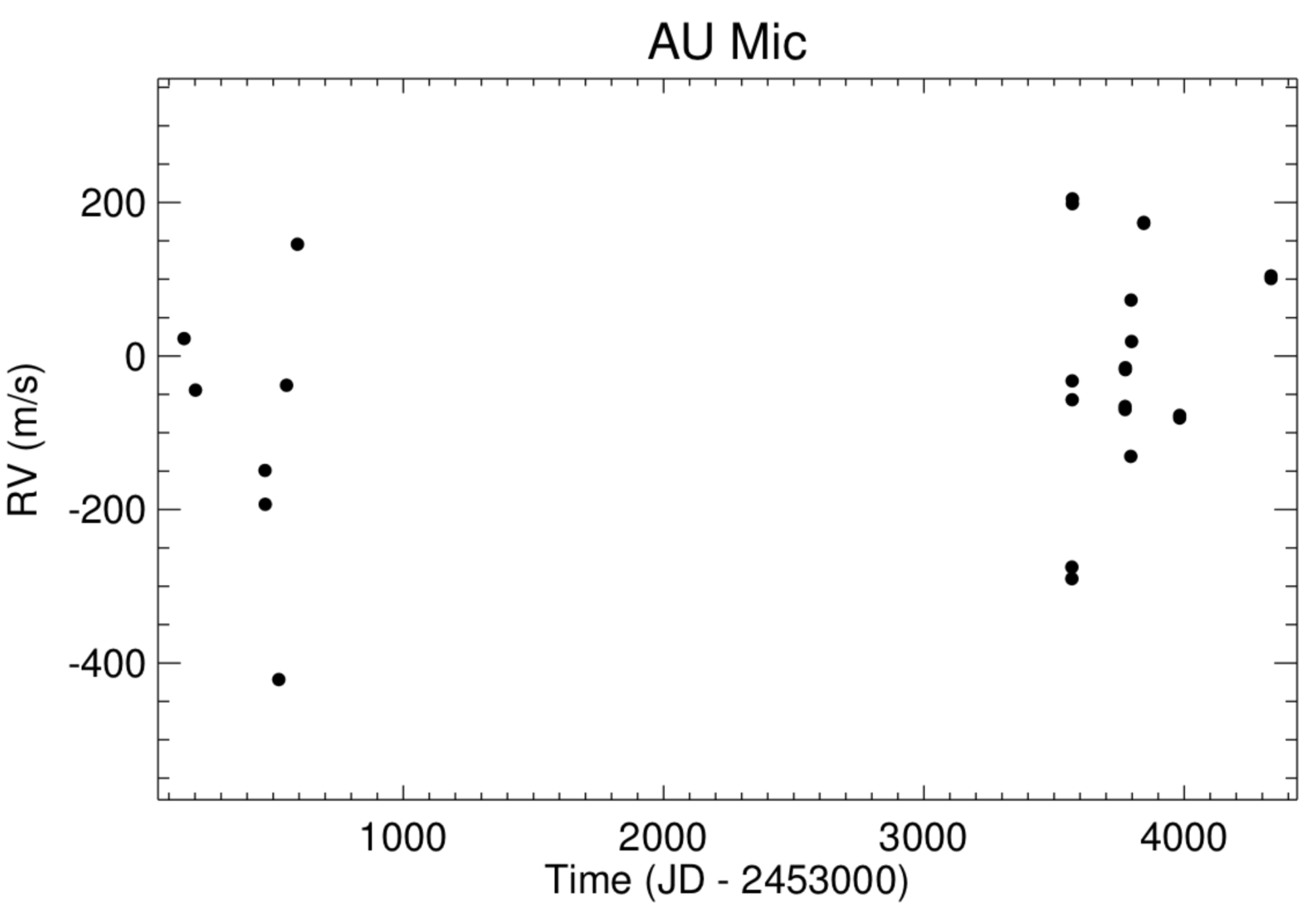}
    \includegraphics[width=90mm]{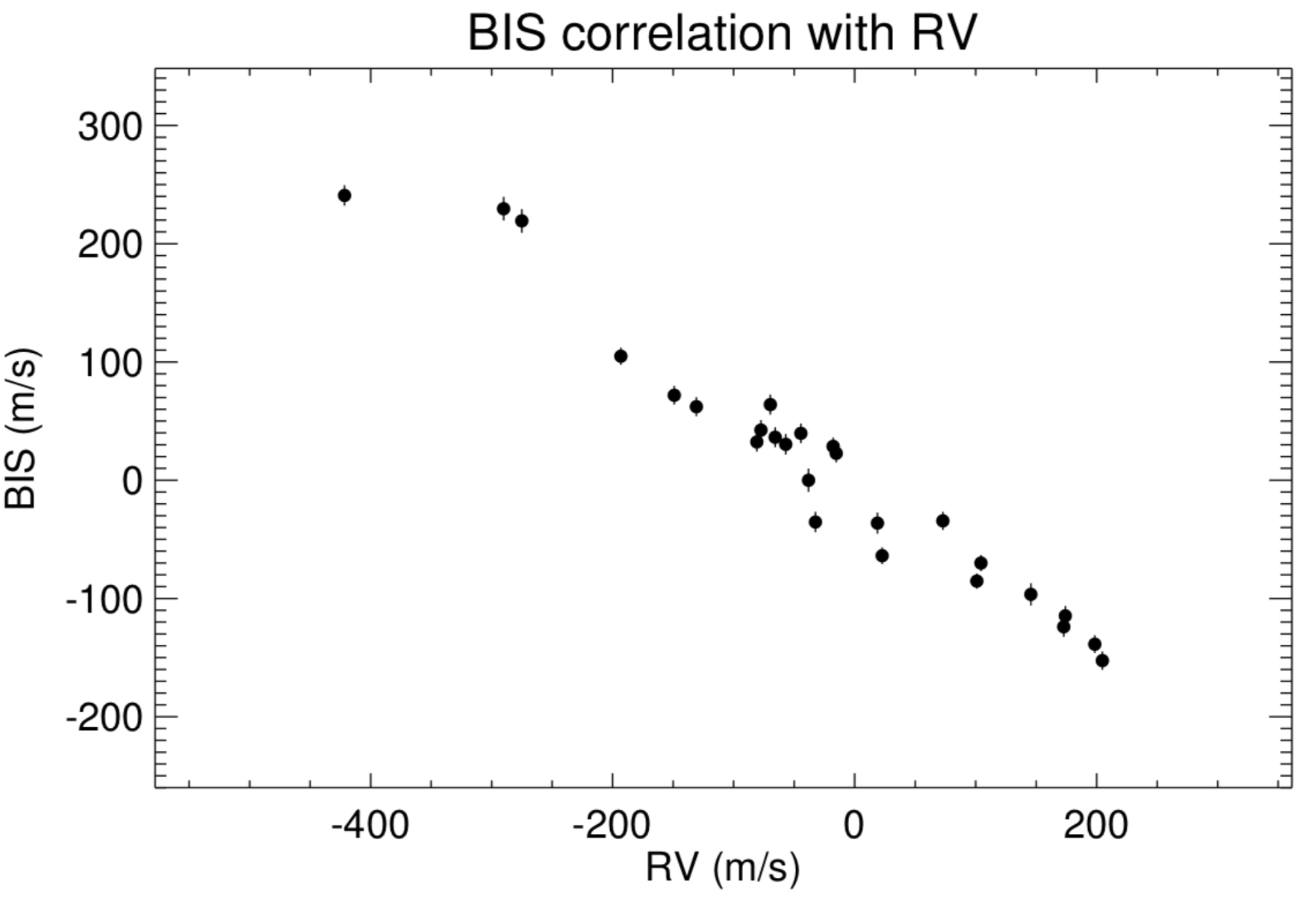}\\
    \includegraphics[width=90mm]{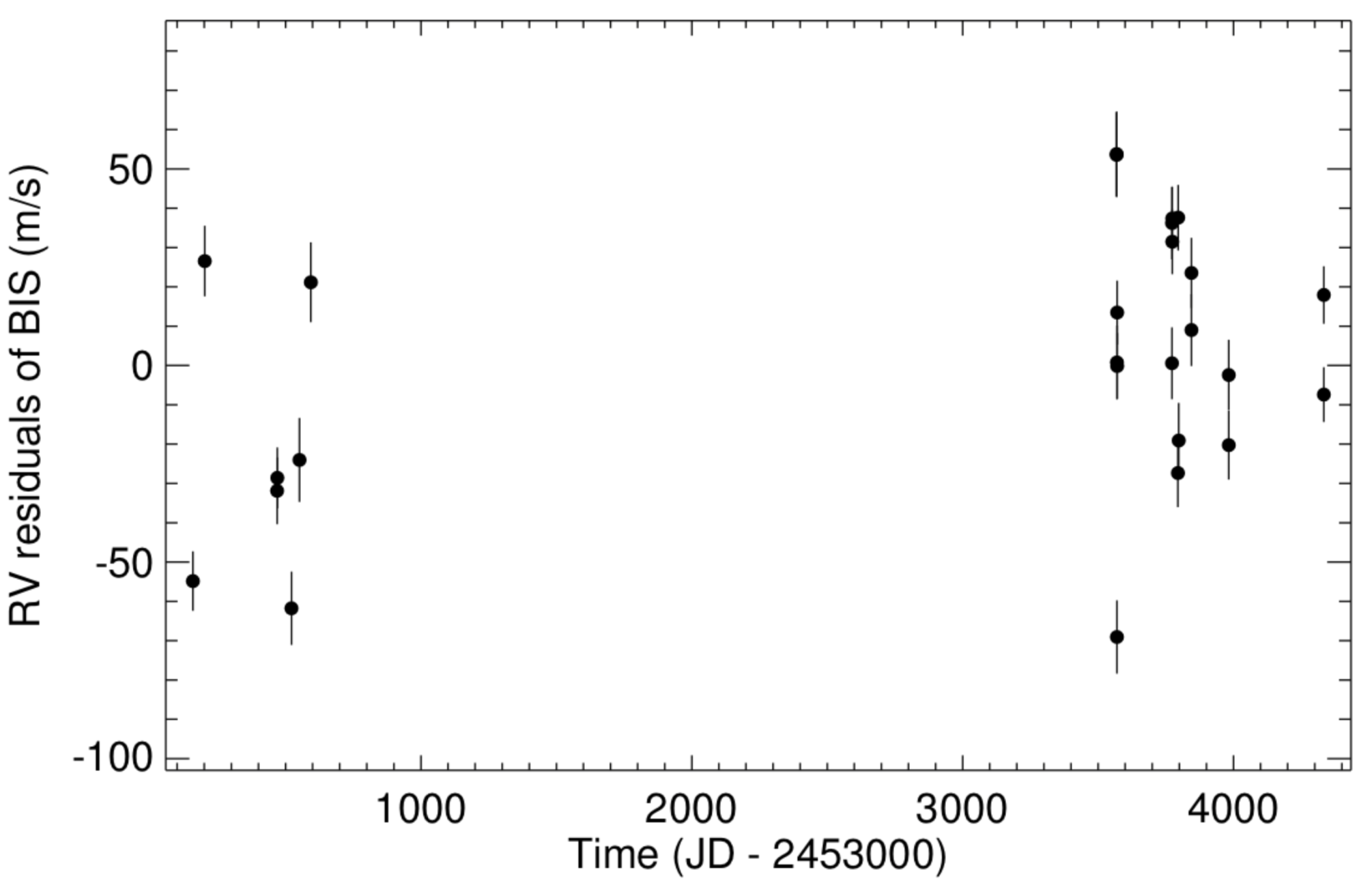}
    \end{tabular}
      \caption{AU~Mic initial RV time series (top left), correlation of RV data with the bisector velocity span (top right), and RV corrected from the bisector velocity span correlation (bottom).}
  \label{RV_curve}
\end{figure*}

AU~Mic has been observed 10 times by NaCo/ESO between 2004 and 2010, in L' band (see Tab.~\ref{obs}), and 26 times by HARPS/ESO over 11 years (from 2004 to 2015). HARPS provides high precision RV data (<1m/s). The data were reduced by the instrument data reduction software. Setups of simultaneous Thorium-Argon exposures were used. Thanks to its mass, its close distance (10~pc) and its long RV time base data (11~years), AU~Mic is a good test for MESS2, as it is one of the few cases available for which the parameter space probed by RV and DI significantly overlap. The time sampling of the RV data is not homogeneous: a first set of measurements has been taken between 2004 and 2005, and a second set of data was obtained 8~years later. AU~Mic is a very active M1V star with a rotation period P = 4.5~d, a $v\times\sin i=9.3~km~s^{-1}$ \citep{Torres.2006}, and peak-to-peak V-band light curve amplitudes up to $\Delta$V = 0.10~mag \citep{Messina.2010}. Its RV are strongly impacted by the presence of long lived spots/plages.  Yet, the RV can be very efficiently corrected thanks to a very good correlation (Pearson's coefficient $\approx$-0.97) between the RV and bisector \citep[see Fig.~\ref{RV_curve}, and other examples of correction of the RV time series using the correlation with the bissector span in][]{Lagrange.2013}. Basically, the corrected $RV_{corr}$ are $RV_{obs}-\alpha BIS$, where $\alpha$ is the slope of the [BIS, RV] data. This correction decreases the RV RMS from 156.6~m/s to 33.9~m/s. We then use the RV corrected time series shown in Fig.~\ref{RV_curve} (bottom). The RV measurements and bisector velocity spans are given in Table~\ref{tab_RV}.

\subsection{Simulations Setup} 
We set the semi-major axis and mass ranges to respectively [0.05;130]~au and [0.05,80]~$M_{Jup}$. The semi-major axis step is logarithmic between 0.05 and 2~au, then linear between 2 and 130~au with a smaller step between 2 and 8~au. We chose this composite sampling to better probe the closer-in region where the RV data are more sensitive. We choose a uniform distribution for the eccentricity between 0 and 0.6, in order to test the detectability of weakly to moderately eccentric planets that would more likely maintain the debris disk dynamically stable. We also choose the inclination range to be [83$^{\circ}$,97$^{\circ}$] assuming then that the planets orbit within the plane of the edge-on disk, with the possibility to generate planets just above or underneath the disk plane within 7$^{\circ}$, following the example of the planets of our Solar System. We use $N_{gen}=10000$.

\begin{table}
\centering
\tiny
\caption{Periods of observation for the AU~Mic DI data.}
\begin{tabular}{c c c}
\hline\hline
Obs. period & run & number of observations\\ [0.5ex]
\hline
June, July, Sept. 2004 & 073.C-0834(A) & 7\\
July 2010 & 085.C-0675(A) & 1\\
Sept. 2010 & 085.C-0277(B) & 1\\ [1ex]
\end{tabular}
\label{obs}
\end{table}

\subsection{Combining the DI data}
\begin{figure}[!h]
  \centering
    \includegraphics[width=90mm]{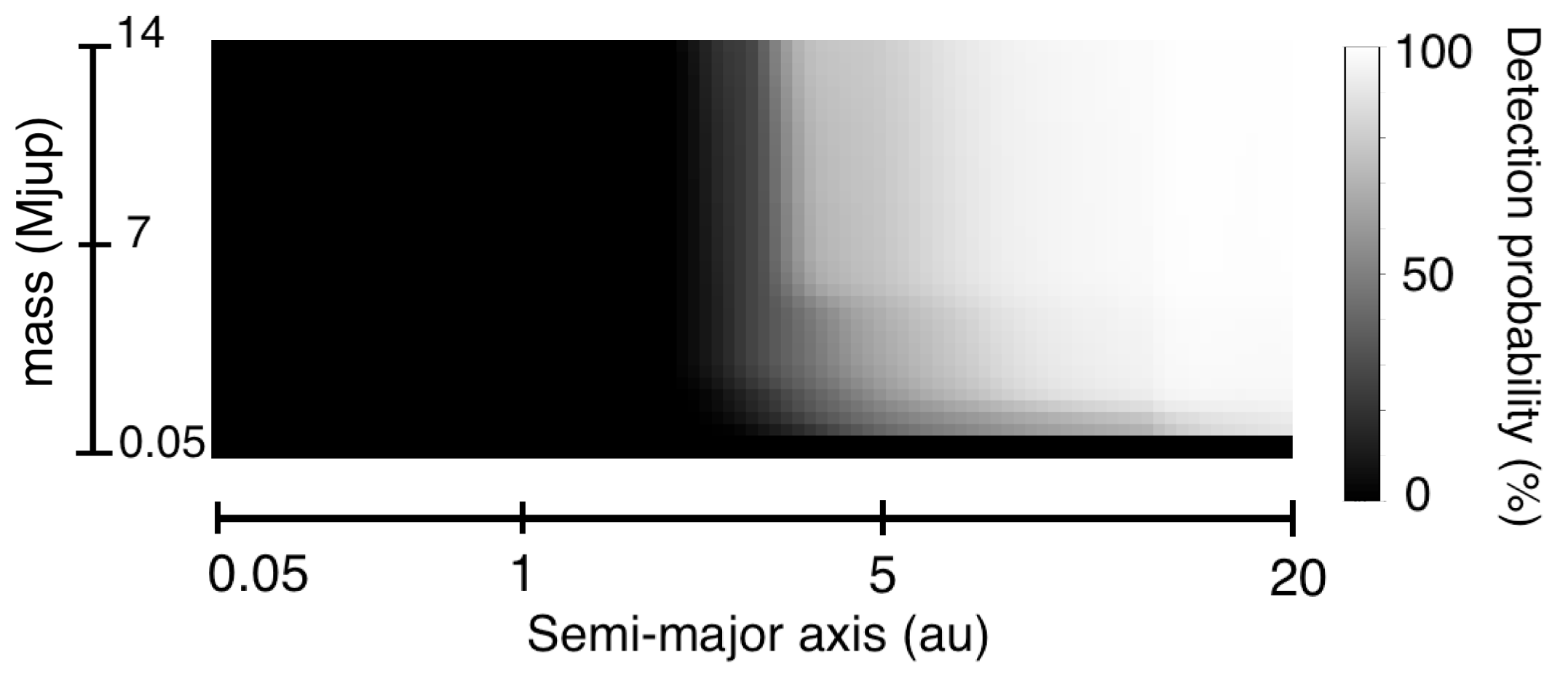} \\
    \includegraphics[width=90mm]{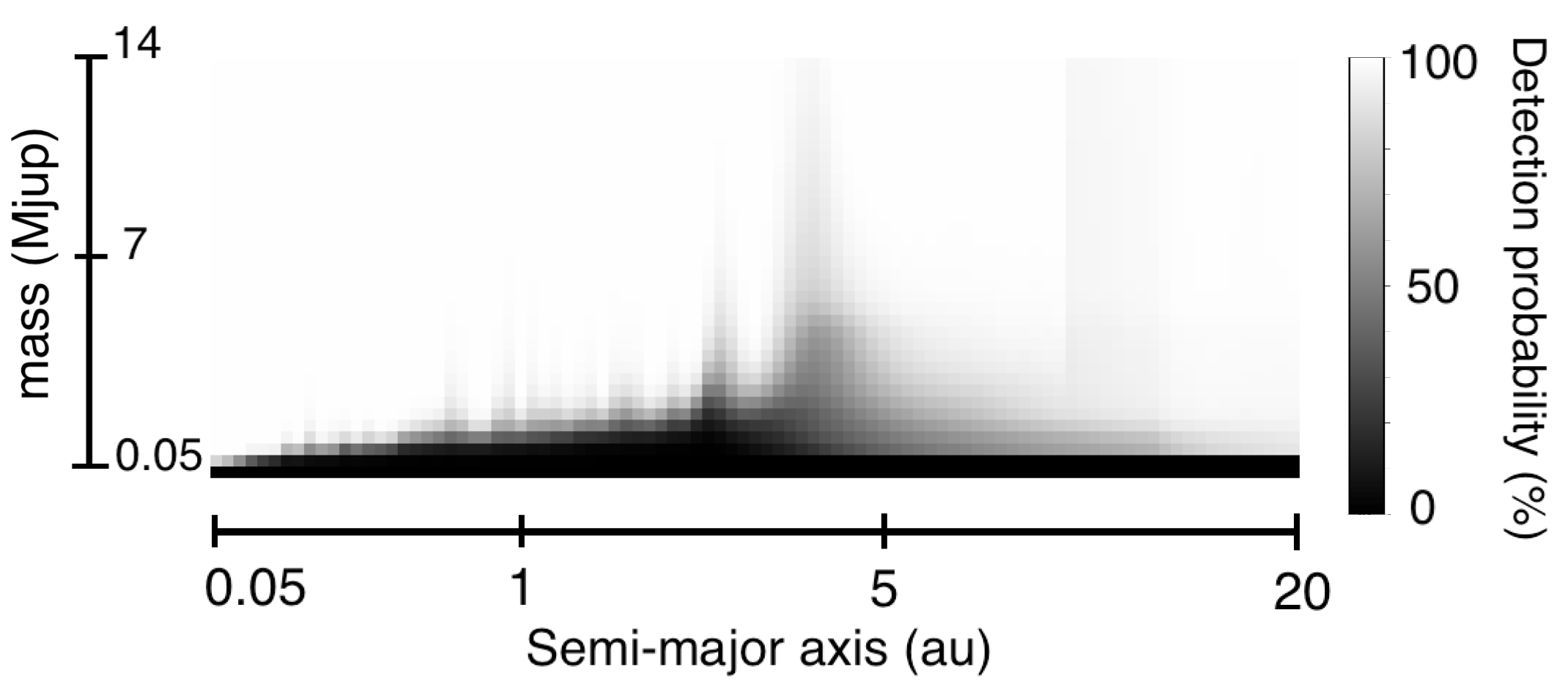} \\
    \includegraphics[width=90mm]{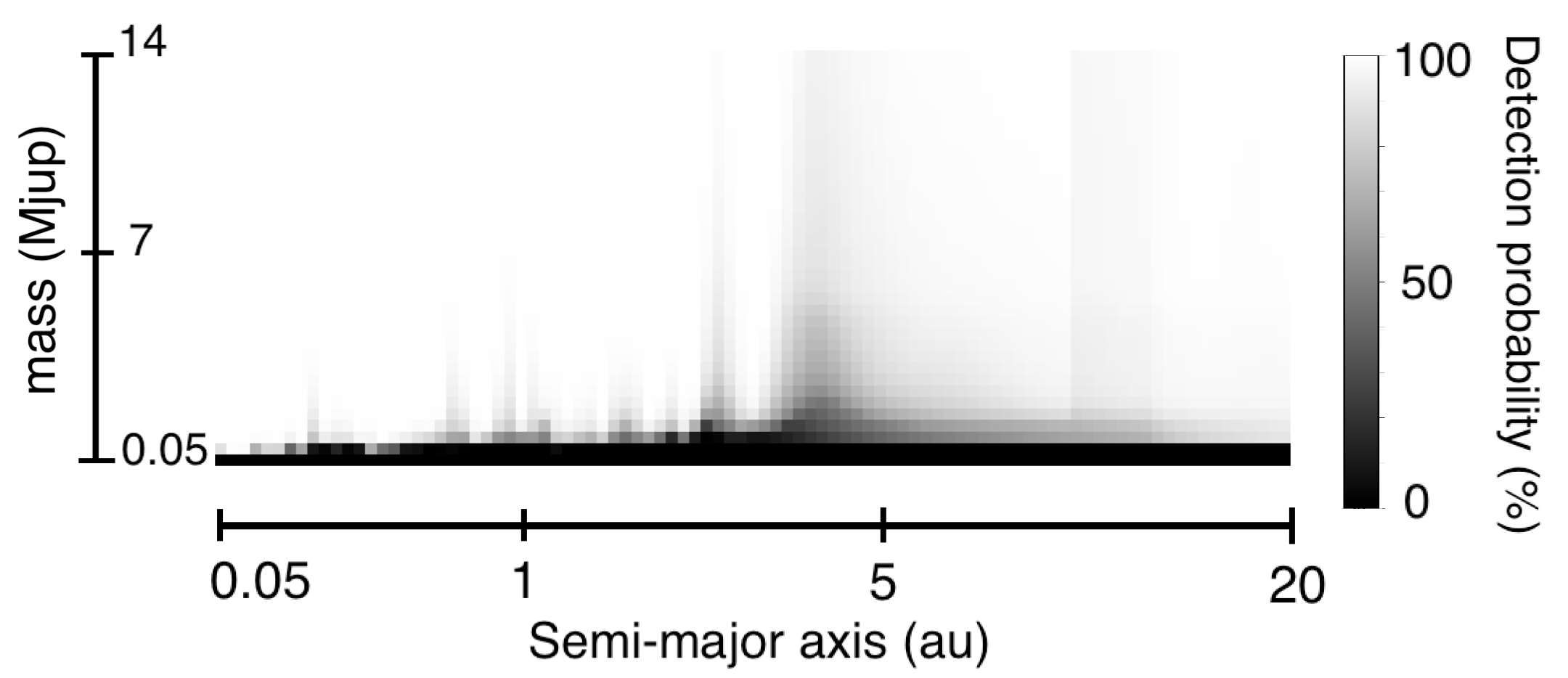}
  \caption{Detection probabilities when using all AU~Mic DI data (top), all DI and RV data with our RMS approach (middle), and all DI and RV data with the LPA approach (bottom). We use a planet mass range of [0.5,14]~M$_{Jup}$, a separation range of [0.05,20]~au, an eccentricity range of [0,0.6], an inclination range of [83$^{\circ}$,97$^{\circ}$], and $N_{gen}=10000$. }
\label{DI+RV}
\end{figure}

\begin{figure*}[!h]
       \begin{minipage}[c]{.50\linewidth}
      \includegraphics[width=95mm,trim=2cm 12cm 2cm 2.5cm,clip]{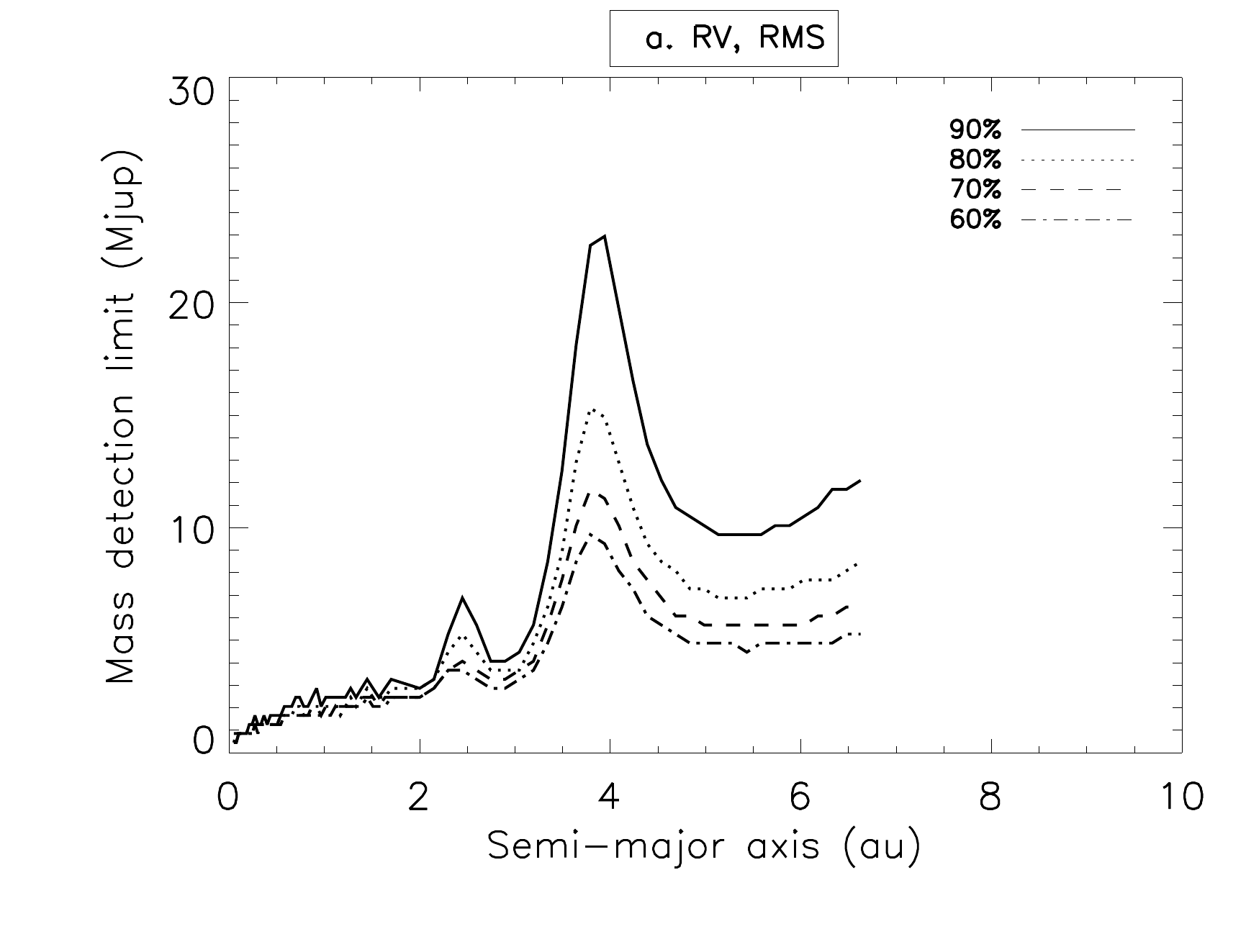}       
   \end{minipage} \hfill   
        \begin{minipage}[c]{.50\linewidth}
      \includegraphics[width=95mm,trim=2cm 12cm 2cm 2.5cm,clip]{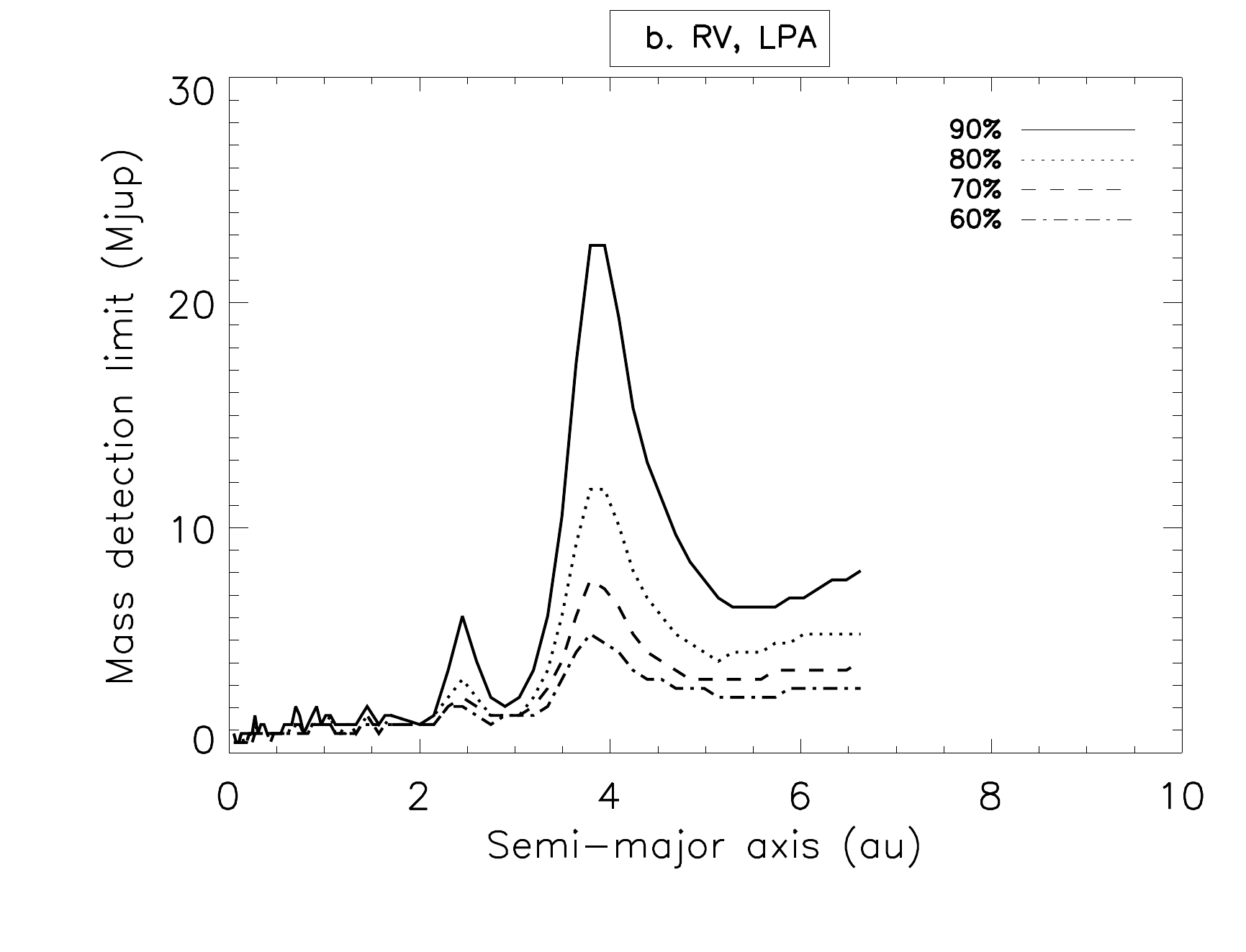}        
   \end{minipage} \hfill   
        \begin{minipage}[c]{.50\linewidth}
      \includegraphics[width=95mm,trim=2cm 12cm 2cm 2.5cm,clip]{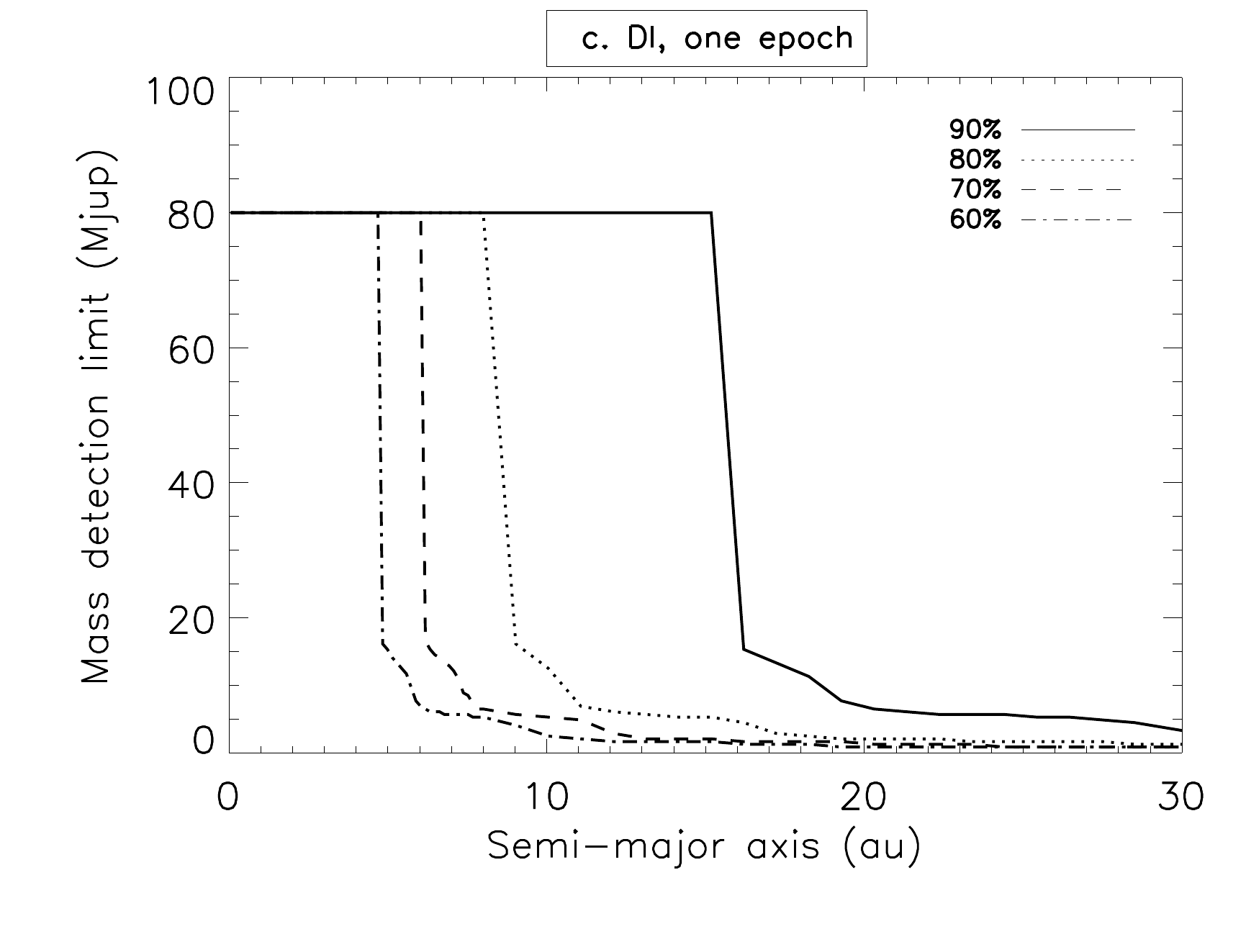}        
   \end{minipage} \hfill  
        \begin{minipage}[c]{.50\linewidth}
      \includegraphics[width=95mm,trim=2cm 12cm 2cm 2.5cm,clip]{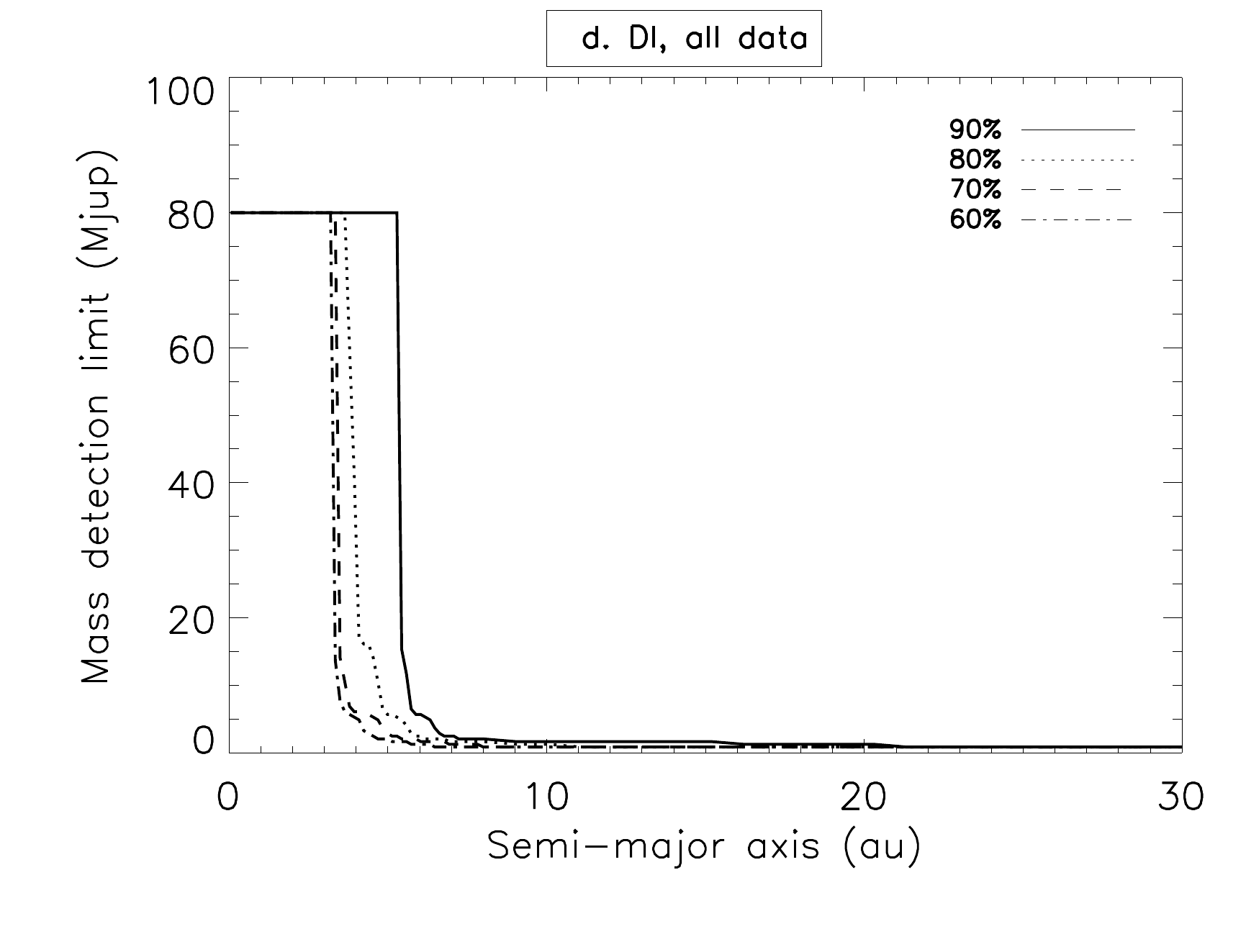}
     	 \end{minipage} \hfill   
        \begin{minipage}[c]{.50\linewidth}
      \includegraphics[width=95mm,trim=2cm 12cm 2cm 2.5cm,clip]{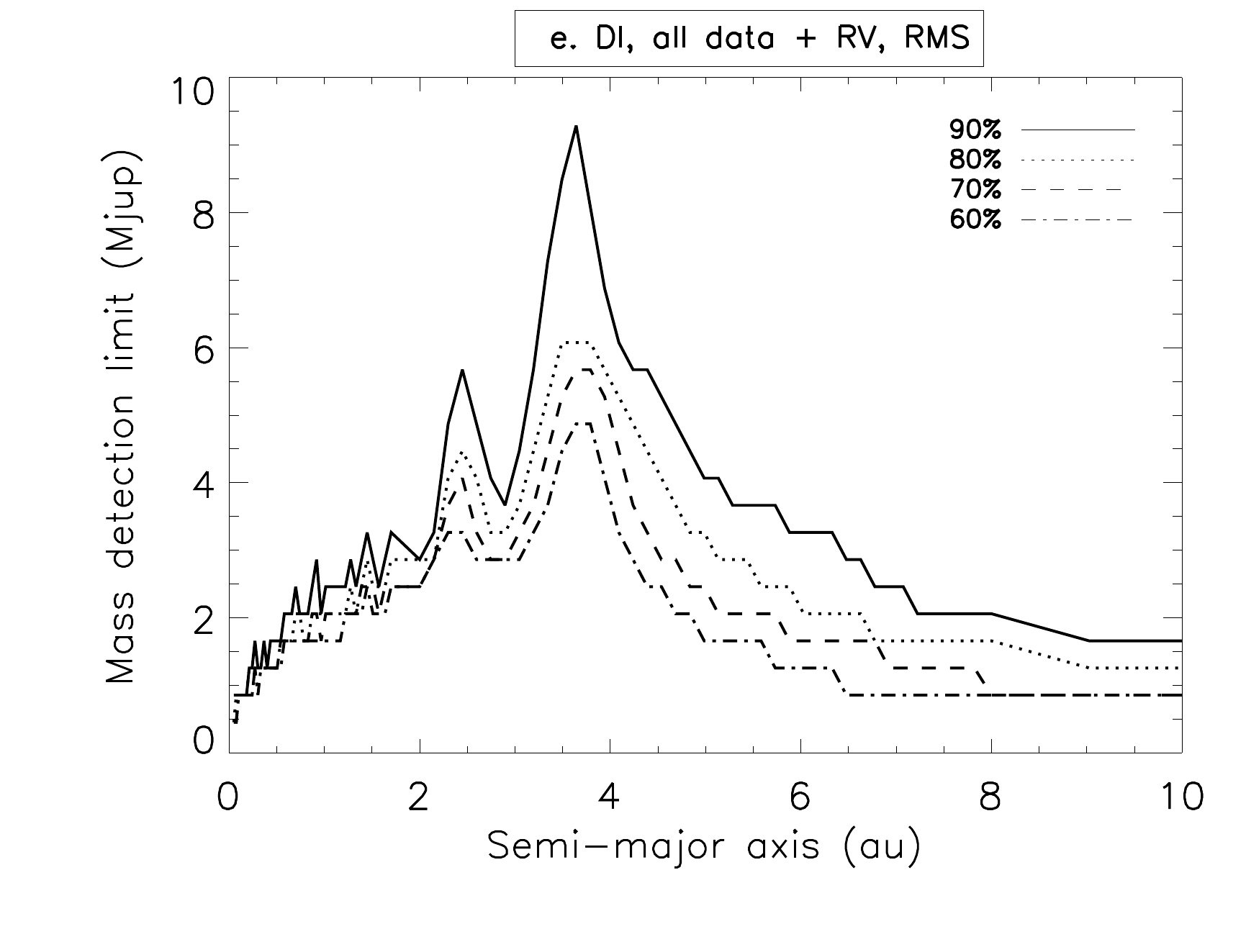}
   \end{minipage} \hfill  
        \begin{minipage}[c]{.50\linewidth}
      \includegraphics[width=95mm,trim=2cm 12cm 2cm 2.5cm,clip]{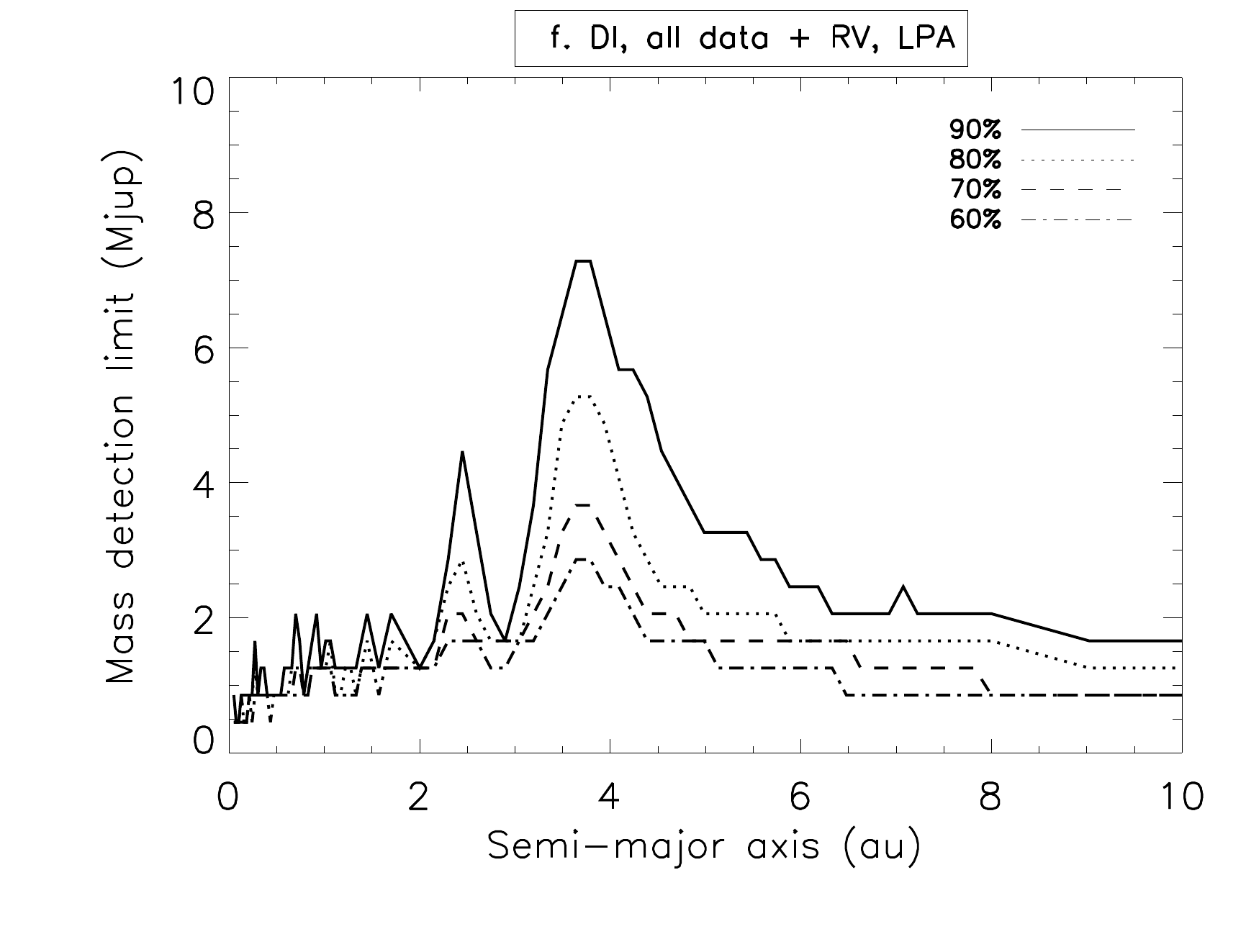}
     	 \end{minipage} \hfill 
    \caption{Mass detection limits for AU~Mic (for detection probabilities of 60\%, 70\%, 80\%, 90\%) when using RV measurements solely with RMS (a.) and LPA (b.) approaches, one DI data from one epoch (c.) and all DI data (d.), and finally both DI and RV data using the RMS (e.) and LPA (f.) approaches. We use an eccentricity range of [0,0.6], an inclination range of [83$^{\circ}$,97$^{\circ}$], and $N_{gen}=10000$. Note: the peaks around 2.7 and 4.3~au are due to the imperfect temporal sampling of the RV data. The low mass detection limits of less than 1~M$_{Jup}$ obtained at large separations are in fact upper limits, corresponding to the lowest values provided by the evolution models by \citet{Baraffe.2003} and BT-Settl model atmosphere \citet{Allard.2012}. }
    \label{isoproba}
    \end{figure*}

We use the giant planet detection probability maps that we had previously derived for each observation epoch \citep[for further detail, see ][]{Delorme.2012a,Lannier_stat.2016}. Figure~\ref{DI+RV} (top) shows the detection probability map ([0.05,20]~au and [0.5,14]~M$_{Jup}$) that we obtain after following the process described in Section~\ref{methods}. Figure~\ref{isoproba} (middle) illustrates the mass detection limit improvement when using all the available DI data. Note that the low mass detection limits of less than 1~M$_{Jup}$ obtained at large separations are in fact upper limits. Indeed, the evolution models by \citet{Baraffe.2003} and BT-Settl model atmosphere \citet{Allard.2012} do not provide values for masses less than 1~M$_{Jup}$.
\\
Compared to the use of a one-epoch DI data, our detection probabilities are improved at relatively short separations and for low-mass planets, typically for semi-major axis range 3-15~au and masses under 10~M$_{Jup}$. In practice, significant improvements will be obtained if the planet period is $\approx 4$ the time baseline or less, so that the planets move enough on their orbits to be detectable at one epoch. 
\\
As the combination of multiple epochs mostly improves the detection probability for low-mass planets on short separations, in the following we will focus on companions with masses between 0.5 and 14~$M_{Jup}$ and within 20~au.

\subsection{Combining the AU~MIC DI and RV data}
\begin{table*}[!h]
\centering
\caption{AU~Mic initial RV measurements with errors, and bisector velocity span with errors in km.s$^{-1}$, for the corresponding date of observation.}
\begin{tabular}{|c | c | c | c | c |}
\hline
BJD-2453000 & 	RV &	RV errors & bisector velocity span & bisector velocity span errors \\ [0.5ex]
\hline
157.898424  &	0.022713    &	0.002816    &	-0.063852   &	0.007041	\\
201.823450	&   -0.044461   &	0.003343    &	0.039650    &	0.008356	\\
468.892370  &	-0.149082   &	0.003154    &	0.071819    &	0.007885	\\
469.843534	&   -0.193213	&	0.002877	& 	0.104851	&	0.007191	\\
521.894368	&	-0.421558	&	0.003464	&	0.240802	&	0.008661	\\	
551.803998	&	-0.038124	&	0.003970	&	0.000000	&	0.009926	\\	
593.622139	&   0.145622	&	0.003777	&	-0.096523	&	0.009442	\\	
3568.505024	&   -0.290126	&	0.003997	&	0.229599	&	0.009993	\\	
3568.515706	&	-0.275093	&	0.004017	&	0.219286	&	0.010044	\\
3569.494659	&	-0.032413	&	0.003461	&	-0.035353	&	0.008653	\\
3569.505549	&	-0.056961	&	0.003481	&	0.030390	&	0.008703	\\
3570.560455	&	0.198430	&	0.003022	&	-0.138663	&	0.007555	\\
3570.571449	&	0.204715	&	0.003110	&	-0.152526	&	0.007775	\\
3772.914293	&	-0.065683	&	0.003395	&	0.036329	&	0.008488	\\
3772.924248	&	-0.069677	&	0.003422	&	0.063948	&	0.008556	\\
3773.911762	&	-0.017799	&	0.002991	&	0.028629	&	0.007477	\\
3773.926196	&	-0.015262	&	0.003064	&	0.022738	&	0.007661	\\
3794.882288	&	-0.130789	&	0.003228	&	0.062227	&	0.008070	\\
3795.885873	&	0.072862	&	0.003100	&	-0.034385	&	0.007749	\\
3797.857541	&	0.018862	&	0.003558	&	-0.036264	&	0.008894	\\
3844.801157	&	0.174065	&	0.003308	&	-0.114677	&	0.008270	\\
3844.811169	&	0.172775	&	0.003392	&	-0.123908	&	0.008479	\\
3982.534763	&	-0.080822	&	0.003252	&	0.032365	&	0.008129	\\
3982.544392	&	-0.077453	&	0.003327	&	0.042436	&	0.008318	\\
4333.530396	&	0.100939	&	0.002592	&	-0.085290	&	0.006480	\\
4333.541066	&	0.104374	&	0.002721	&	-0.070044	&	0.006802	\\ [1ex]
\hline
\end{tabular}
\label{tab_RV}
\end{table*}
\begin{figure*}[!h]
  \centering
  \begin{tabular}{cc}
    \includegraphics[width=95mm,trim=2cm 12cm 2cm 2.5cm,clip]{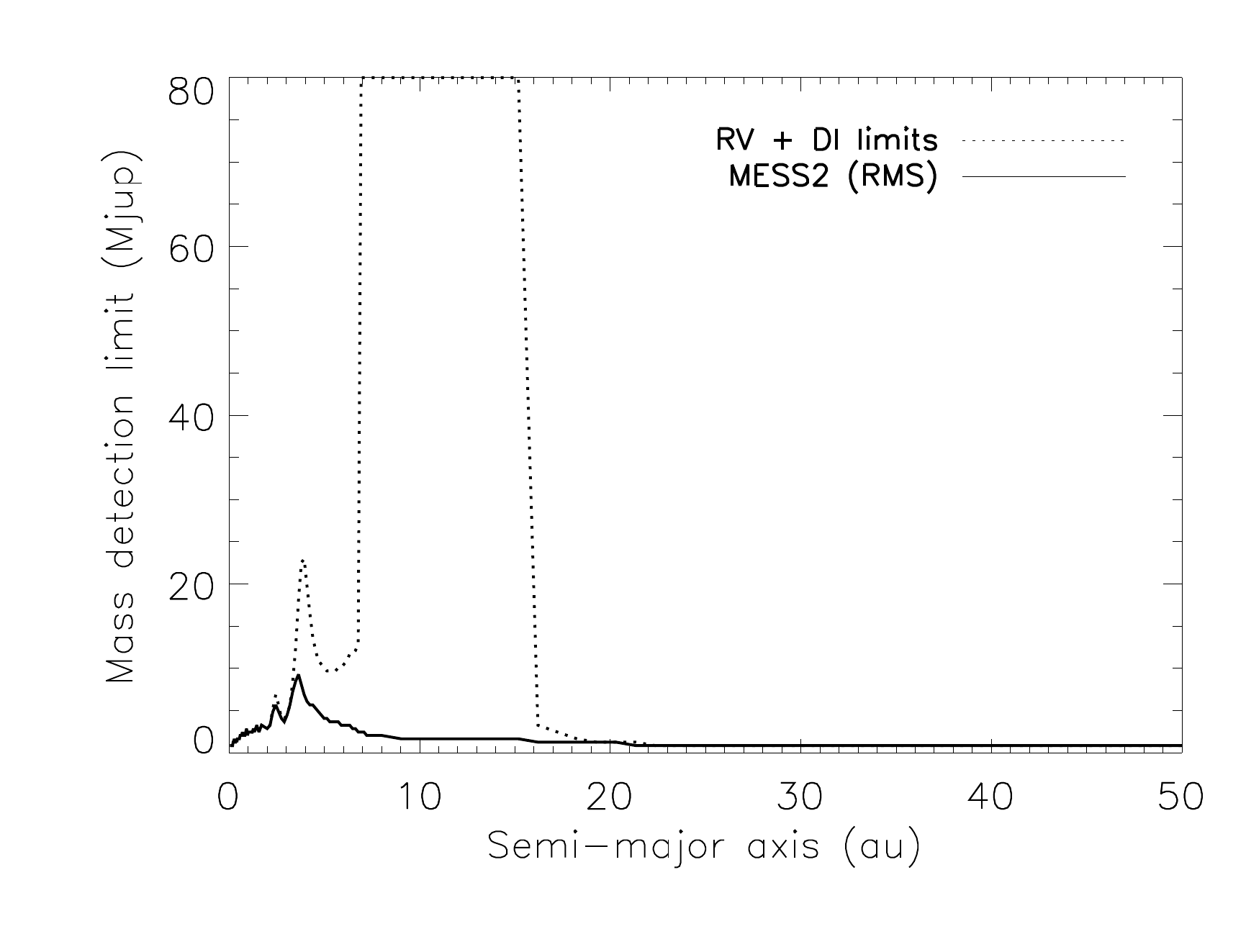}
    \includegraphics[width=95mm,trim=2cm 12cm 2cm 2.5cm,clip]{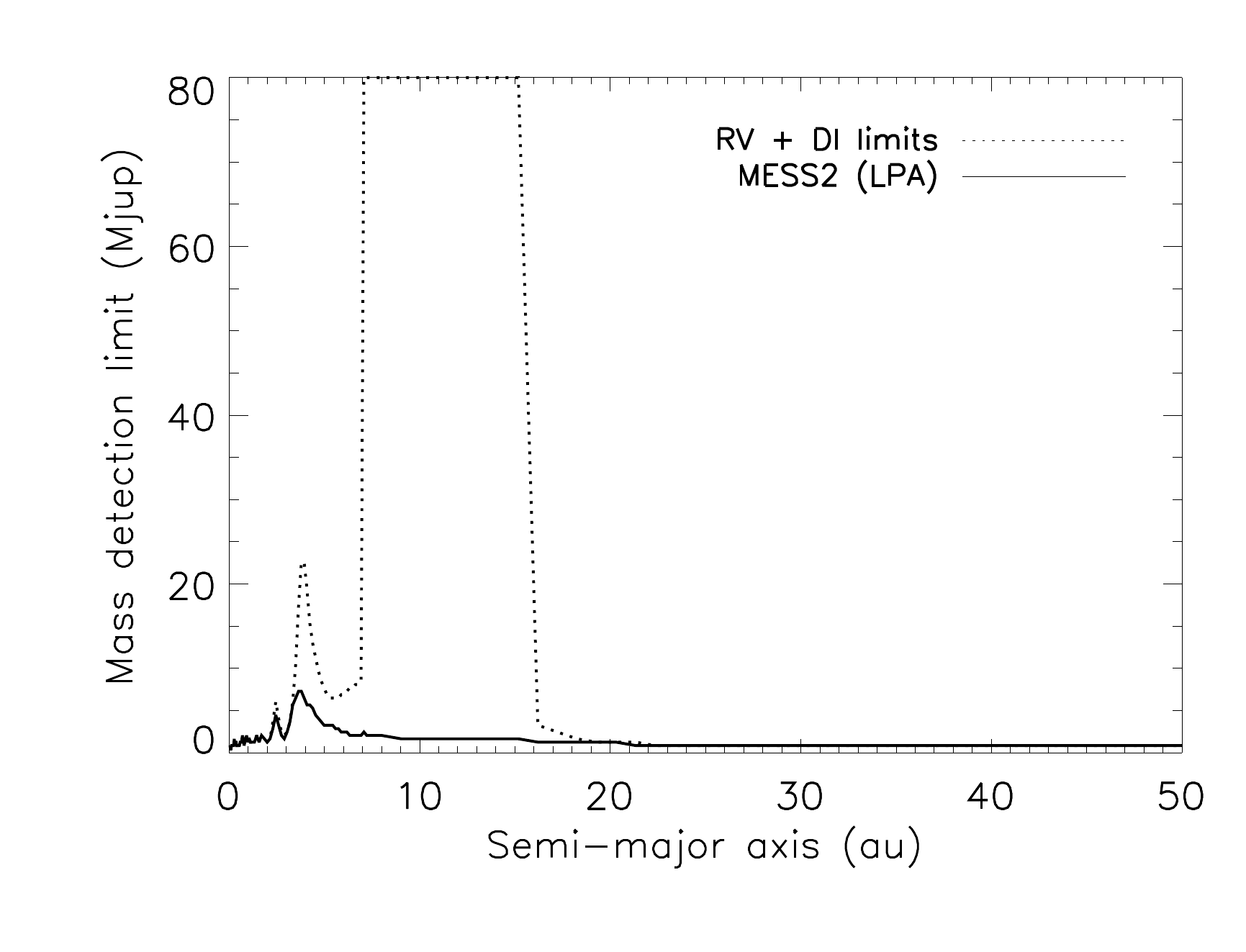}\\
    \includegraphics[width=95mm,trim=2cm 12cm 2cm 2.5cm,clip]{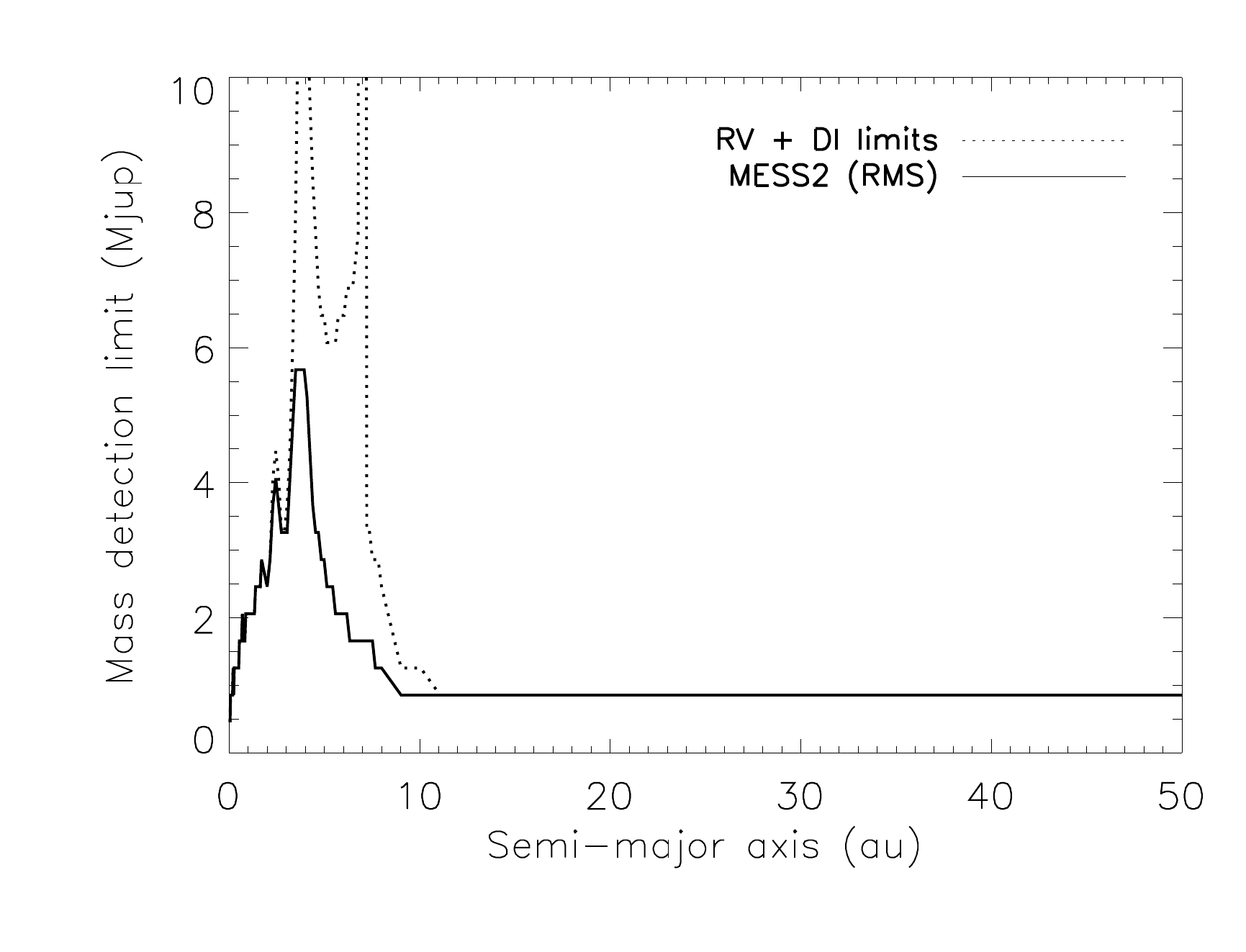}
    \includegraphics[width=95mm,trim=2cm 12cm 2cm 2.5cm,clip]{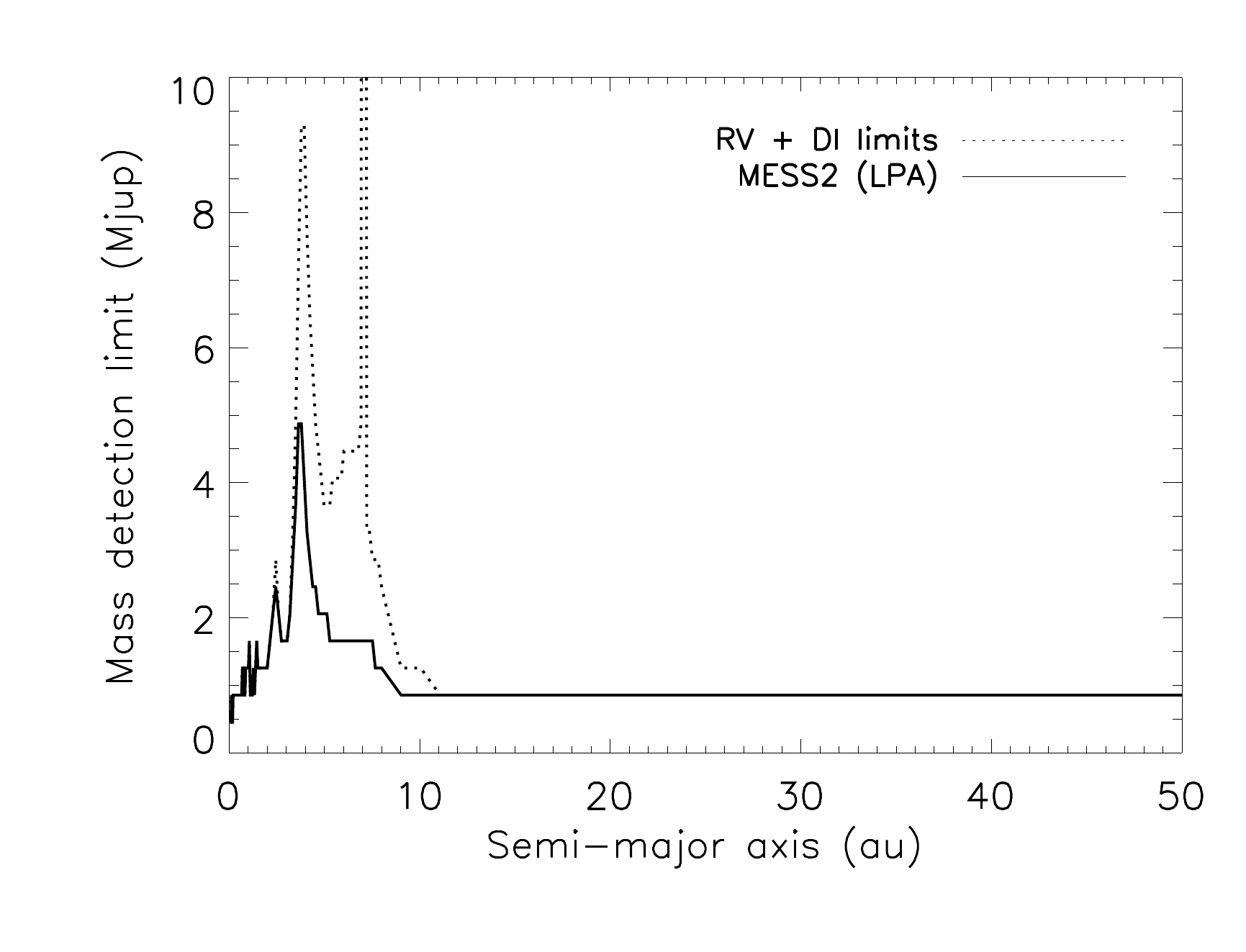}\\
  \end{tabular}
  \caption{Comparison of the detection limits (90\% top, 75\% bottom) when combining the RV and DI independent detection limit curves, and when the RV and DI data are combined in a common analysis through MESS2 with the RMS (left) and LPA (right) approaches. We use an eccentricity range of [0,0.6], an inclination range of [83$^{\circ}$,97$^{\circ}$], and $N_{gen}=10000$.}
  \label{MESS2_MESS}
\end{figure*}
\begin{figure}[!h]
  \centering
    \includegraphics[width=90mm,trim=2cm 12.5cm 2cm 2.5cm,clip]{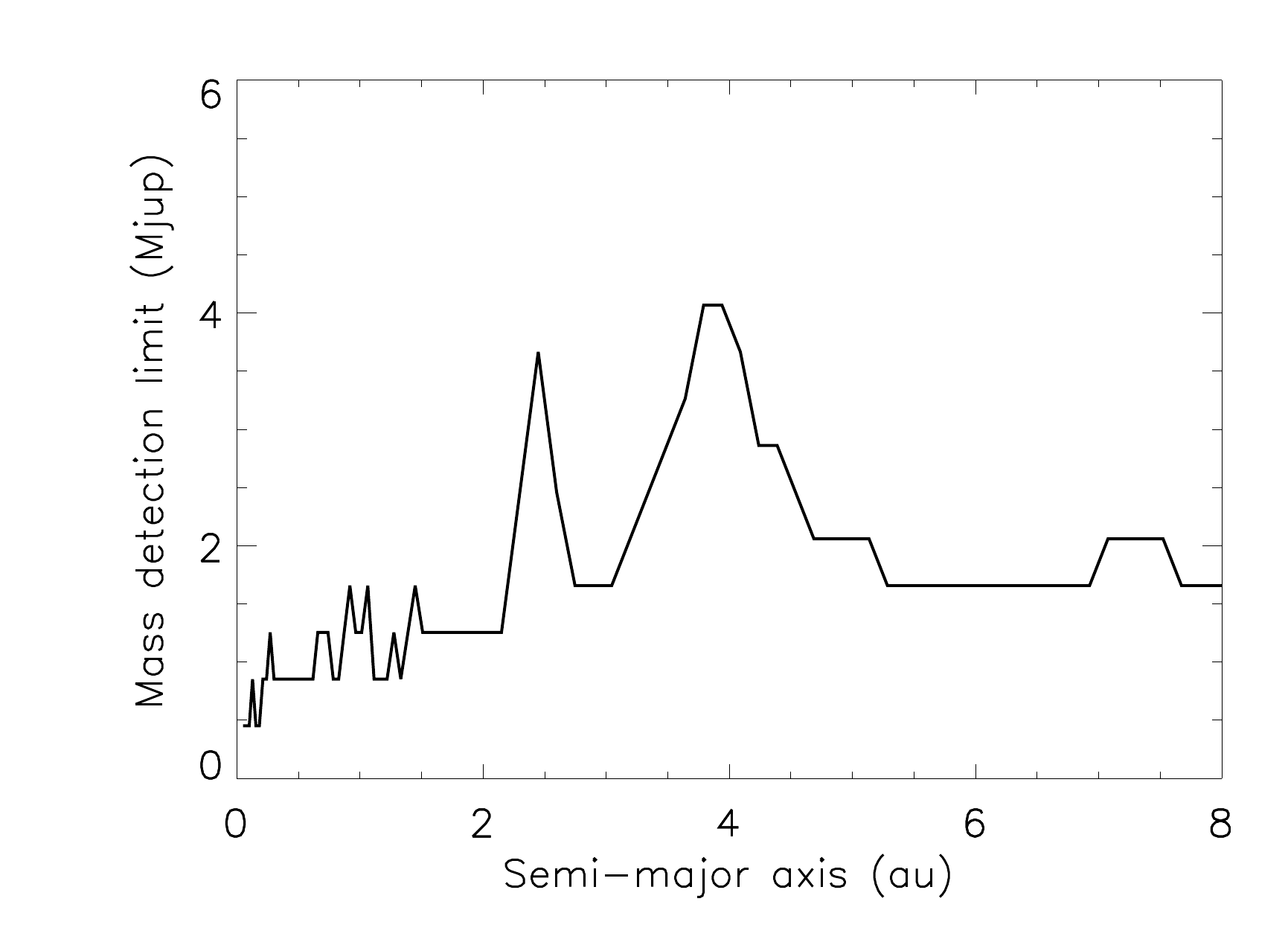}    
    \caption{Mass detection limits (90\%) obtained after combining AU~Mic DI and RV data considering circular orbits, with the LPA approach. We use i=[83$^{\circ}$,97$^{\circ}$], and $N_{gen}=10000$.}
  \label{ecc_0}
\end{figure}
Figure~\ref{DI+RV} (middle and bottom) shows the probability maps obtained when combining the RV and DI data, using the RMS and LPA approaches (middle and bottom, respectively). Figure~\ref{isoproba} shows the associated detection limits. Figure~\ref{MESS2_MESS} shows the comparison of the best detection limits that we could obtain with MESS2 (solid line) and without (dashed line). For the calculation of the detection limits represented by the dashed line, we compute the detection limits obtained with RV data on the one hand and with the most sensitive DI dataset on the other hand, taking the best mass limit between the two of them for each probed separation (in the text we call this approach "the standard combination", while the "self-consistent combination" refers to the DI and RV data combination within MESS2).
\\
As expected, combining self-consistently DI and RV data rather than using the standard combination, dramatically improves the detection limits at shorter separations compared to the use of a one-epoch DI data only, since this parameter range is very well investigated by RV but not yet reachable by the DI instrumental sensitivity. On the contrary, using RV data does not improve the detection of wider-orbit planets, since the RV technique is not sensitive to large separations.
\\
Second, we find that the combination of both RV and DI observations significantly improves the detection limits up to 20~au. For instance, at 5~au, without MESS2, we find a detection limit at 90\% probability of $\approx$8~M$_{Jup}$ using the LPA approach only, while MESS2 provides a mass limit down to $\approx$3.5~M$_{Jup}$ at this separation (LPA method). When an overlap between the RV and DI data is possible, using a standard DI and RV independent detection limits combination becomes obsolete, because this standard combination under-estimates the number of generated planets that could be detected, resulting in worse detection limits. This has a direct impact on the planet frequency derived from surveys, since the planet occurrence rate was therefore systematically overestimated within the parameter domain overlapped by DI and RV data. MESS2 gives more robust detection limits and probabilities in this parameter range: for instance, with the LPA approach, twice as many synthetic planets of 2~M$_{Jup}$ at 4~au are detected than using the standard combination of DI and RV detection limits. The difference between using MESS2 instead of a standard combination of RV and DI detection limits is higher for smaller planets close to the detection limits that are usually used to define the best sensitivity of a data set. At larger separations, there is still an important difference also close to the detection limits, that is due to the use of multi-epoch DI data (see Fig.~\ref{MESS2_MESS}). 
\\
\\
Note that for separations typically between 7 and 16~au, the poor detection limits at 90\% probability derived from the standard combination (see Fig.~\ref{MESS2_MESS}) are due to projection effects, even bright companions would be hidden behind the star during more than 10\% of their orbit.
\\
Note also that we show in Fig.~\ref{ecc_0} the detection limits that we obtain by testing circular orbits instead of eccentric ones, as the debris disk around AU~Mic can prevent the presence of too eccentric, wide-orbit planets. In the following, we still use the eccentricity range [0,0.6].



\subsection{Scheduling future DI observations.}
\begin{figure*}
\begin{center}
\begin{tabular}{cc}
      \includegraphics[width = 90mm]{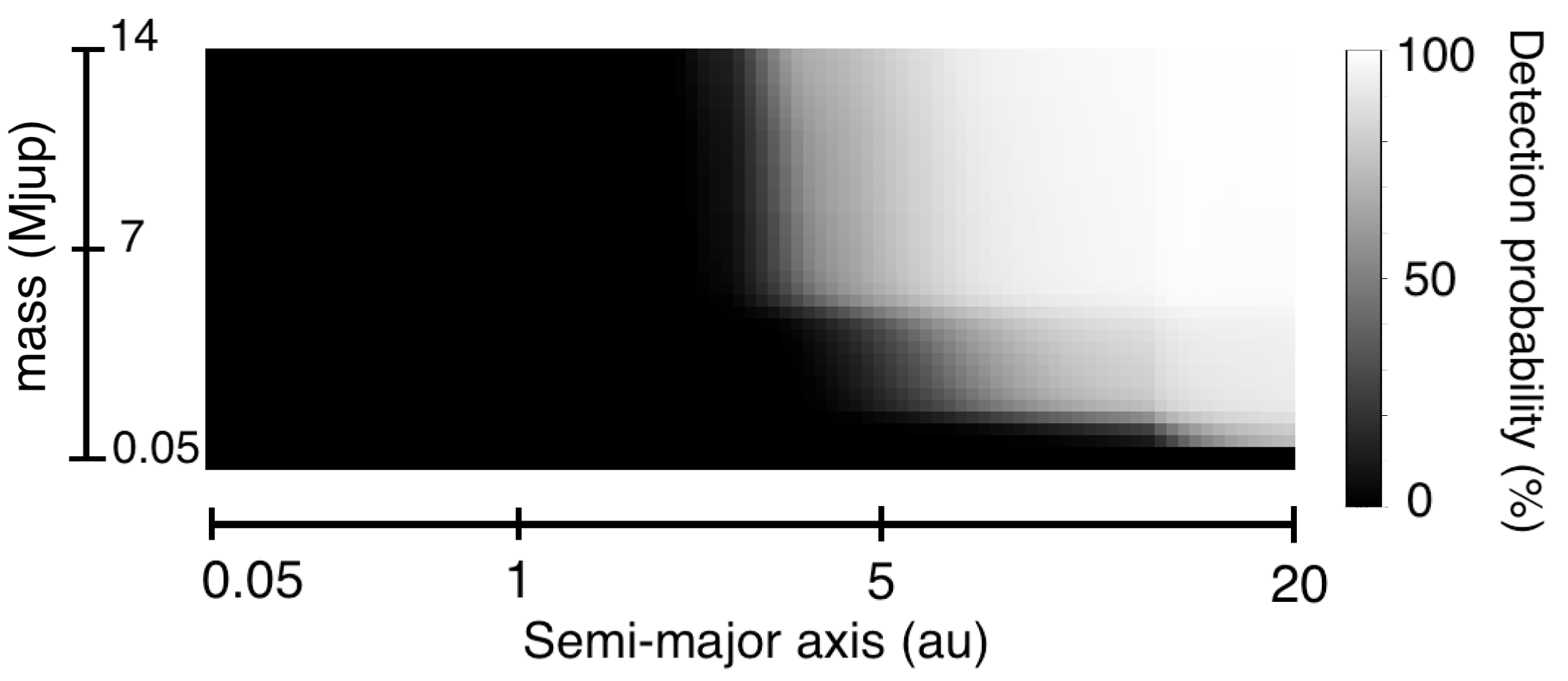}
      \includegraphics[width = 90mm]{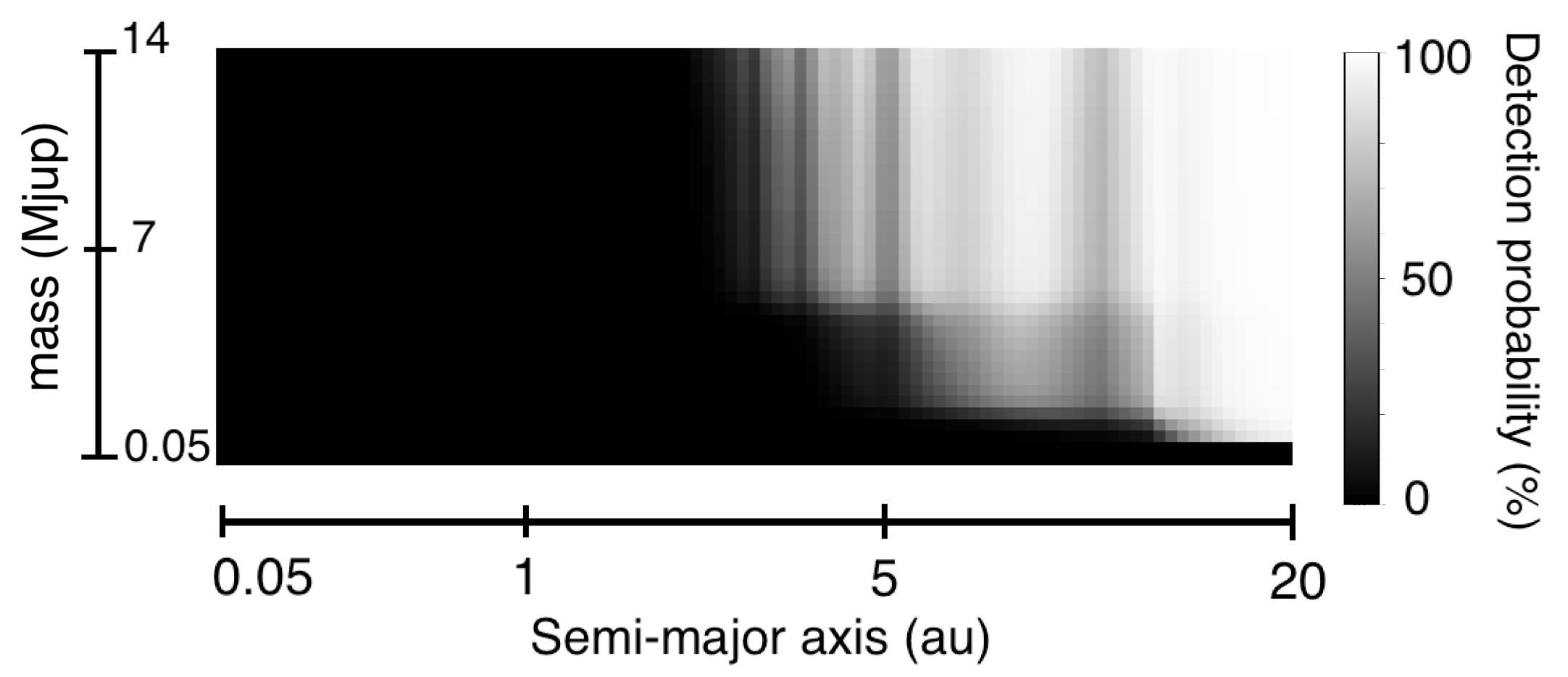}
\end{tabular}
   \caption{Simulation of the combination of two DI observations. The second DI data is assumed to be obtained 5 (left) and 25~years (right) later. We use a planet mass range of [0.5,14]~M$_{Jup}$, a separation range of [0.05,20]~au, an eccentricity range of [0,0.6], an inclination range of [83$^{\circ}$,97$^{\circ}$], and $N_{gen}=10000$.}
   \label{single}
\end{center}
\end{figure*}

\begin{figure*}[h]
\begin{center}
     \includegraphics[width = 150mm,trim=2cm 13cm 2cm 3cm,clip]{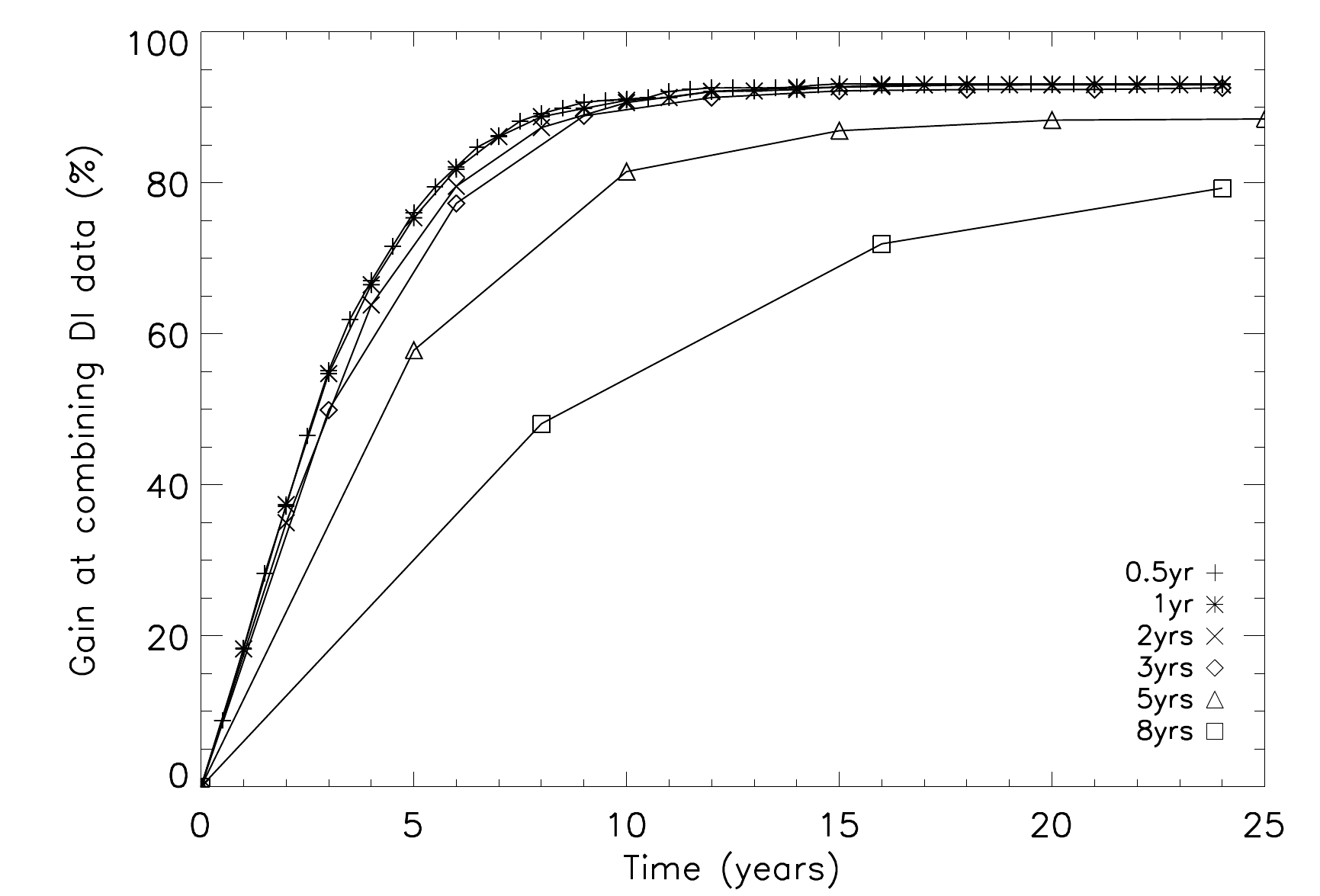}
\caption{Gain in percentage at combining DI data separated by a given time lapse, over a given time baseline. The mass range is 0.5-14~M$_{Jup}$ and semi-major axis range is 4-8~au. We use an eccentricity range of [0,0.6], an inclination range of [83$^{\circ}$,97$^{\circ}$], and $N_{gen}=10000$.}
\label{interval_strategy}
\end{center}
\end{figure*}

Another independent use of MESS2 is the optimisation of the schedule of new DI observations, more precisely to predict the best time intervals and time baselines to obtain new DI data of a star, depending on the semi-major axis and mass ranges. 
\\
In order to illustrate this feature of the code we constructed several sets of simulated observations, by combining two copies of the same DI observation (we use the first epoch observation of AU~Mic, taken on 2004-06-10) and changing the time span between the two simulated data. This allowed us to assess the changes in detection probability as a function of the time between the single epochs. 
 Figure.~\ref{single} shows the probability maps when a second DI data is taken 5 and 25 years later. Note that we see poor-gain vertical lines that move towards larger semi-major axis when the time lapse increases. They are associated to the revolution periods of the planets: for instance, planets located at 2.5~au and 7.3~au cover their entire orbit respectively in 5 and 25~years, so there is no benefit to re-observe them after a full revolution. Figure~\ref{interval_strategy} illustrates the detection probability gain obtained when combining different DI data separated by 0.5, 1, 2, 3, 5 or 8~years, over different time baselines, for planet between 0.5 and 14~M$_{Jup}$ located in the restrictive semi-major axis range [4-8]~au. The gain is defined as the median value of the detection probability after the combination of several DI data, over the median value of the detection probability using a one-epoch data, within the planet masses and semi-major axes restrictions. Of course, observing a star as often as possible gives the best gains, but a very high observational frequency is not necessary: observing only three times AU~Mic over ten years gives almost the same gain than observing the target more frequently, if we search for planets with masses 0.5-14~M$_{Jup}$ at 4-8~au, with appropriate time sampling. Indeed, the plateau that we observe around an improvement of $\approx$90\% at observing the star several times, is reachable with a 10-year time baseline, for DI observations separated by less than 5 years.

\section{Concluding remarks and perspectives}
\label{conclusion}
\begin{figure*}
  \centering
    \includegraphics[width=130mm,trim=2cm 13cm 2cm 3cm,clip]{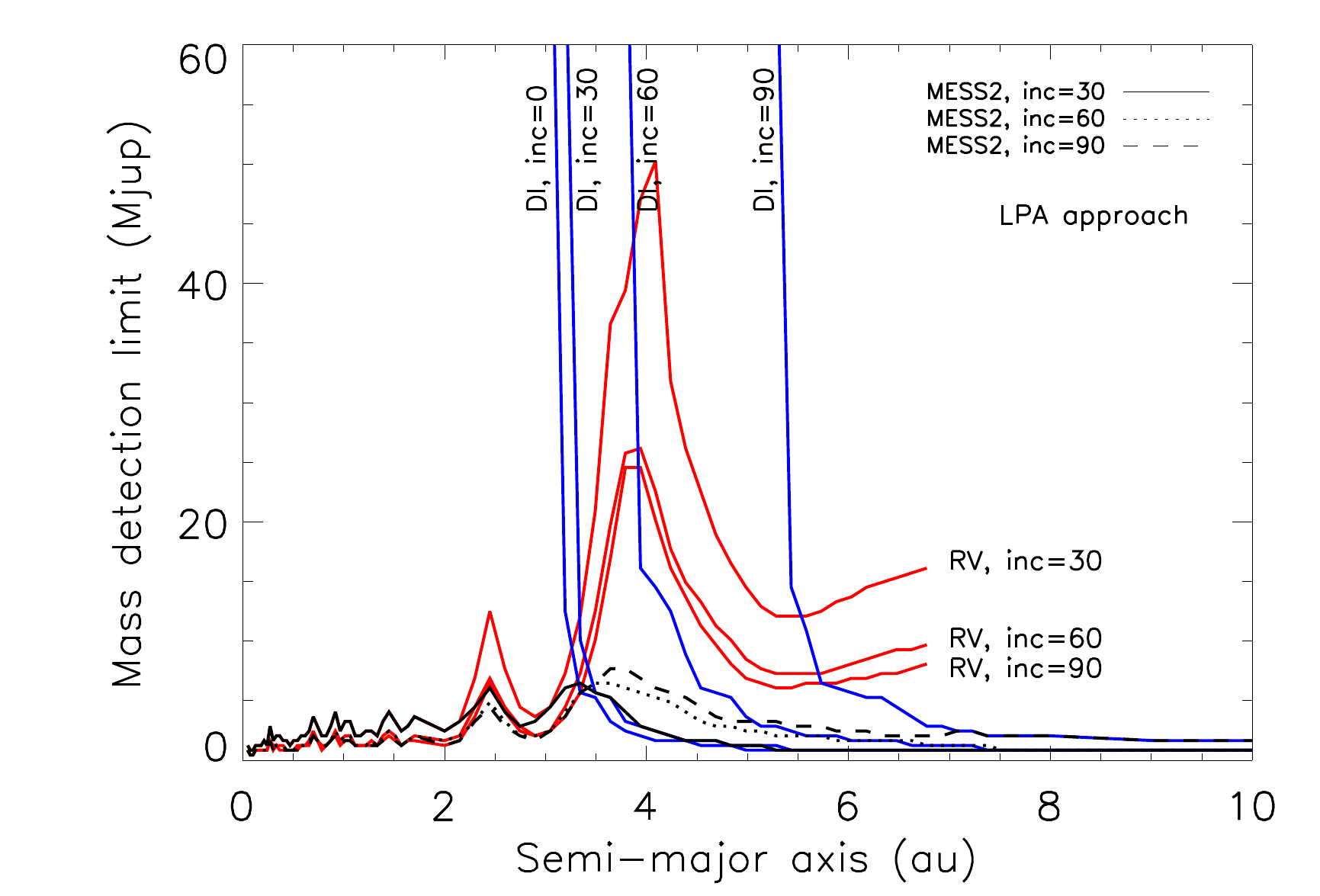}
    \caption{Mass detection limits (90\%) obtained combining all AU~Mic DI data with MESS2 on the one hand (blue lines), and using MESS2 on RV data with the LPA approach on the other hand (red lines), for different orbital inclinations. The detection limits derived by MESS2 (multi-epoch DI and RV data combination) are represented by the black lines. We use an eccentricity range of [0,0.6], and $N_{gen}=10000$.}
  \label{inclination_limdet}
\end{figure*}

We presented the MESS2 code (Multi-epochs multi-purposes exoplanet simulation system), a statistical tool that combines multi-technique data. Our aim with MESS2 was 1) to properly constrain giant planet populations from the inner to outer regions of stars (using RV and DI techniques) to eventually constrain the models of planet population synthesis, and 2) to optimise the scheduling of DI observations. A MESS2 application is presented in this paper in the case of the late-type star AU~Mic. Another application will be presented in the case of a pulsating early type star, $\beta$~Pictoris, in a coming paper (Lagrange et al. 2016, in prep.).
\\
In the case of AU~Mic, located at 10~pc, we could bridge the gap between RV and DI data and constrain the presence of giant planets with masses $\ge$2~M$_{Jup}$ (using probabilities of 60\%) from a fraction of au to several tens of AU, and $\le$1~M$_{Jup}$ outside the 2-4~au range. The detection limits obtained after the self-consistent combination of RV and DI data with MESS2 are significantly better than those obtained with the standard combination of RV and DI independent detection limits. In particular, between 7 and 16~au, we showed that more than 90\% of planets down to 2~M$_{Jup}$ are detectable. On the opposite, the standard combination of RV and DI independent detection limits cannot exclude with a 90\% probability the presence of brown dwarfs of any mass at these separations. In addition to the fact that MESS2 provides more constrained and more robust detection limits, not using a self-consistent tool such as MESS2 for AU~Mic leads to overestimate the planet occurence rate. Additional RV data and/or future GAIA astrometric data as described by \citet{Sozzetti.2014} could better constrain the intermediate separations, i.e. less than a few astronomical units. 
\\
The detection probabilities computed depend on the configuration of the planets orbits (i.e. the inclination with respect to the line of sight, and the position angle of the orbit with respect to the North). First, the RV and DI methods are sensitive in opposite ways to the planets inclinations. If the system is observed edge-on (as AU~Mic), then combining sets of DI and RV data together greatly improves the probability to detect planets at all separations. On the contrary, if the system is observed pole-on, then the RV measurements do not constrain the detection probabilities. Moreover, if one assumes circular and pole-on orbits, there will be no gain at combining all DI data. The intermediate cases between those extremes are more common, like HR8799 \citep[25$^\circ$ with respect to the plane of the sky,][]{Contro.2015}, TW~Hydrae \citep[7$^\circ$ for the outer disk and 4.3$^\circ$ for the inner disk,][]{Setiawan.2008,Pontoppidan.2008} or HD141569 \citep[51$^\circ$,][]{vanderPlas.2015}. Fig.~\ref{inclination_limdet} shows that the different inclinations affect differently the detection limits. Second, it is also possible, within the MESS2 code, to take into account the position angle of the system by constraining the longitude of ascending mode. Constraining this orbital parameter is relevant in cases where a structure such as a circumstellar disk is detected in the considered DI data and affects the data sensitivity. Taking into consideration the position angle of the system does not affect the detectability of RV planets, while the impact on the detectability of DI planets can be strong. Applying MESS2 on $\beta$~Pictoris data is a good test for constraining the position angle of the orbits of the synthesized planets (see Lagrange et al. 2016, in prep.). 
\\
The impact of each detection technique also depends on the stellar properties. For instance, the younger the stars are, the more sensitive DI is, but this is not necessary true for RV (it depends on the stellar activity, pulsation and rotation levels). 
\\
\\
Finally, the combination of RV and DI data when RV trends are observed can be used to constraint the orbital and physical parameters of the potential planet that would produce this trend. Indeed, either the planet is detected within the DI data and thus its parameters are constrained, either it is not imaged and then the DI detection limits can be used to put upper limits on its physical properties. This application of MESS2 will be the object of a future paper \citep{Bonavita.2016}.

\begin{acknowledgements}
We thank the staff of ESO-VLT and ESO-La Silla for their support at the telescope. We acknowledge support from the French National Research Agency (ANR) through the GuEPARD project grant ANR10-BLANC0504-01 and GIPSE project grant ANR-14-CE33-0018. We thank Pascal Rubini for his work on the software SAFIR.
\end{acknowledgements}

\bibliographystyle{aa}
\bibliography{biball}

\begin{thebibliography}{42}
\expandafter\ifx\csname natexlab\endcsname\relax\def\natexlab#1{#1}\fi

\bibitem[{{Allard} {et~al.}(2012){Allard}, {Homeier}, \&
  {Freytag}}]{Allard.2012}
{Allard}, F., {Homeier}, D., \& {Freytag}, B. 2012, Royal Society of London
  Philosophical Transactions Series A, 370, 2765

\bibitem[{{Baraffe} {et~al.}(2003){Baraffe}, {Chabrier}, {Barman}, {Allard}, \&
  {Hauschildt}}]{Baraffe.2003}
{Baraffe}, I., {Chabrier}, G., {Barman}, T.~S., {Allard}, F., \& {Hauschildt},
  P.~H. 2003, \aap, 402, 701

\bibitem[{{Beuzit} {et~al.}(2006){Beuzit}, {Feldt}, {Dohlen}, {Mouillet},
  {Puget}, {Antichi}, {Baruffolo}, {Baudoz}, {Berton}, {Boccaletti},
  {Carbillet}, {Charton}, {Claudi}, {Downing}, {Feautrier}, {Fedrigo}, {Fusco},
  {Gratton}, {Hubin}, {Kasper}, {Langlois}, {Moutou}, {Mugnier}, {Pragt},
  {Rabou}, {Saisse}, {Schmid}, {Stadler}, {Turrato}, {Udry}, {Waters}, \&
  {Wildi}}]{Beuzit.2006}
{Beuzit}, J.-L., {Feldt}, M., {Dohlen}, K., {et~al.} 2006, The Messenger, 125

\bibitem[{{Biller} {et~al.}(2007){Biller}, {Close}, {Masciadri}, {Nielsen},
  {Lenzen}, {Brandner}, {McCarthy}, {Hartung}, {Kellner}, {Mamajek}, {Henning},
  {Miller}, {Kenworthy}, \& {Kulesa}}]{Biller.2007}
{Biller}, B.~A., {Close}, L.~M., {Masciadri}, E., {et~al.} 2007, \apjs, 173,
  143

\bibitem[{{Binks} \& {Jeffries}(2014)}]{Binks.2014}
{Binks}, A.~S. \& {Jeffries}, R.~D. 2014, \mnras, 438, L11

\bibitem[{{Boccaletti} {et~al.}(2015){Boccaletti}, {Thalmann}, {Lagrange},
  {Janson}, {Augereau}, {Schneider}, {Milli}, {Grady}, {Debes}, {Langlois},
  {Mouillet}, {Henning}, {Dominik}, {Maire}, {Beuzit}, {Carson}, {Dohlen},
  {Engler}, {Feldt}, {Fusco}, {Ginski}, {Girard}, {Hines}, {Kasper}, {Mawet},
  {M{\'e}nard}, {Meyer}, {Moutou}, {Olofsson}, {Rodigas}, {Sauvage},
  {Schlieder}, {Schmid}, {Turatto}, {Udry}, {Vakili}, {Vigan}, {Wahhaj}, \&
  {Wisniewski}}]{Boccaletti.2015}
{Boccaletti}, A., {Thalmann}, C., {Lagrange}, A.-M., {et~al.} 2015, \nat, 526,
  230

\bibitem[{{Bonavita} {et~al.}(2012){Bonavita}, {Chauvin}, {Desidera},
  {Gratton}, {Janson}, {Beuzit}, {Kasper}, \& {Mordasini}}]{Bonavita.2011}
{Bonavita}, M., {Chauvin}, G., {Desidera}, S., {et~al.} 2012, \aap, 537, A67

\bibitem[{{Bonavita} \& et~al.(2016)}]{Bonavita.2016}
{Bonavita}, M. \& et~al. 2016, in preparation

\bibitem[{{Bonfils} {et~al.}(2013){Bonfils}, {Delfosse}, {Udry}, {Forveille},
  {Mayor}, {Perrier}, {Bouchy}, {Gillon}, {Lovis}, {Pepe}, {Queloz}, {Santos},
  {S{\'e}gransan}, \& {Bertaux}}]{Bonfils.2011}
{Bonfils}, X., {Delfosse}, X., {Udry}, S., {et~al.} 2013, \aap, 549, A109

\bibitem[{{Bowler} {et~al.}(2015){Bowler}, {Liu}, {Shkolnik}, \&
  {Tamura}}]{Bowler.2015}
{Bowler}, B.~P., {Liu}, M.~C., {Shkolnik}, E.~L., \& {Tamura}, M. 2015, \apjs,
  216, 7

\bibitem[{{Chauvin} {et~al.}(2010){Chauvin}, {Lagrange}, {Bonavita},
  {Zuckerman}, {Dumas}, {Bessell}, {Beuzit}, {Bonnefoy}, {Desidera}, {Farihi},
  {Lowrance}, {Mouillet}, \& {Song}}]{Chauvin.2010}
{Chauvin}, G., {Lagrange}, A.-M., {Bonavita}, M., {et~al.} 2010, \aap, 509, A52

\bibitem[{{Contro} {et~al.}(2015){Contro}, {Wittenmyer}, {Horner}, \&
  {Marshall}}]{Contro.2015}
{Contro}, B., {Wittenmyer}, R.~A., {Horner}, J., \& {Marshall}, J.~P. 2015,
  Origins of Life and Evolution of the Biosphere, 45, 41

\bibitem[{{Cumming} {et~al.}(2008){Cumming}, {Butler}, {Marcy}, {Vogt},
  {Wright}, \& {Fischer}}]{Cumming.2008}
{Cumming}, A., {Butler}, R.~P., {Marcy}, G.~W., {et~al.} 2008, \pasp, 120, 531

\bibitem[{{Delorme} {et~al.}(2012){Delorme}, {Lagrange}, {Chauvin}, {Bonavita},
  {Lacour}, {Bonnefoy}, {Ehrenreich}, \& {Beust}}]{Delorme.2012a}
{Delorme}, P., {Lagrange}, A.~M., {Chauvin}, G., {et~al.} 2012, \aap, 539, A72

\bibitem[{{Johnson}(2008)}]{Johnson.2008}
{Johnson}, J.~A. 2008, in Astronomical Society of the Pacific Conference
  Series, Vol. 398, Extreme Solar Systems, ed. D.~{Fischer}, F.~A. {Rasio},
  S.~E. {Thorsett}, \& A.~{Wolszczan}, 59

\bibitem[{{Johnson} {et~al.}(2010){Johnson}, {Aller}, {Howard}, \&
  {Crepp}}]{Johnson.2010}
{Johnson}, J.~A., {Aller}, K.~M., {Howard}, A.~W., \& {Crepp}, J.~R. 2010,
  \pasp, 122, 905

\bibitem[{{Kalas} {et~al.}(2004){Kalas}, {Liu}, \& {Matthews}}]{Kalas.2004}
{Kalas}, P., {Liu}, M.~C., \& {Matthews}, B.~C. 2004, Science, 303, 1990

\bibitem[{{Krist} {et~al.}(2005){Krist}, {Ardila}, {Golimowski}, {Clampin},
  {Ford}, {Illingworth}, {Hartig}, {Bartko}, {Ben{\'{\i}}tez}, {Blakeslee},
  {Bouwens}, {Bradley}, {Broadhurst}, {Brown}, {Burrows}, {Cheng}, {Cross},
  {Demarco}, {Feldman}, {Franx}, {Goto}, {Gronwall}, {Holden}, {Homeier},
  {Infante}, {Kimble}, {Lesser}, {Martel}, {Mei}, {Menanteau}, {Meurer},
  {Miley}, {Motta}, {Postman}, {Rosati}, {Sirianni}, {Sparks}, {Tran},
  {Tsvetanov}, {White}, \& {Zheng}}]{Krist.2005}
{Krist}, J.~E., {Ardila}, D.~R., {Golimowski}, D.~A., {et~al.} 2005, \aj, 129,
  1008

\bibitem[{{Lafreni{\`e}re} {et~al.}(2007){Lafreni{\`e}re}, {Marois}, {Doyon},
  {Nadeau}, \& {Artigau}}]{Lafreniere.2007}
{Lafreni{\`e}re}, D., {Marois}, C., {Doyon}, R., {Nadeau}, D., \& {Artigau},
  {\'E}. 2007, \apj, 660, 770

\bibitem[{{Lagrange} {et~al.}(2013){Lagrange}, {Meunier}, {Chauvin}, {Sterzik},
  {Galland}, {Lo Curto}, {Rameau}, \& {Sosnowska}}]{Lagrange.2013}
{Lagrange}, A.-M., {Meunier}, N., {Chauvin}, G., {et~al.} 2013, \aap, 559, A83

\bibitem[{{Lannier} {et~al.}(2016){Lannier}, {Delorme}, {Lagrange}, {Borgniet},
  {Rameau}, {Schlieder}, {Gagn\'{e}}, {Bonavita}, {Malo}, {Chauvin},
  {Bonnefoy}, \& {Girard}}]{Lannier_stat.2016}
{Lannier}, J., {Delorme}, P., {Lagrange}, A.-M., {et~al.} 2016, submitted to
  A\&A

\bibitem[{{Liu}(2004)}]{Liu.2004_nat}
{Liu}, M.~C. 2004, Science, 305, 1442

\bibitem[{{Liu} {et~al.}(2004){Liu}, {Matthews}, {Williams}, \&
  {Kalas}}]{Liu.2004}
{Liu}, M.~C., {Matthews}, B.~C., {Williams}, J.~P., \& {Kalas}, P.~G. 2004,
  \apj, 608, 526

\bibitem[{{Lomb}(1976)}]{Lomb.1976}
{Lomb}, N.~R. 1976, \apss, 39, 447

\bibitem[{{Macintosh} {et~al.}(2014){Macintosh}, {Graham}, {Ingraham},
  {Konopacky}, {Marois}, {Perrin}, {Poyneer}, {Bauman}, {Barman}, {Burrows},
  {Cardwell}, {Chilcote}, {De Rosa}, {Dillon}, {Doyon}, {Dunn}, {Erikson},
  {Fitzgerald}, {Gavel}, {Goodsell}, {Hartung}, {Hibon}, {Kalas}, {Larkin},
  {Maire}, {Marchis}, {Marley}, {McBride}, {Millar-Blanchaer}, {Morzinski},
  {Norton}, {Oppenheimer}, {Palmer}, {Patience}, {Pueyo}, {Rantakyro},
  {Sadakuni}, {Saddlemyer}, {Savransky}, {Serio}, {Soummer},
  {Sivaramakrishnan}, {Song}, {Thomas}, {Wallace}, {Wiktorowicz}, \&
  {Wolff}}]{Macintosh.2014}
{Macintosh}, B., {Graham}, J.~R., {Ingraham}, P., {et~al.} 2014, Proceedings of
  the National Academy of Science, 111, 12661

\bibitem[{{Messina} {et~al.}(2010){Messina}, {Desidera}, {Turatto},
  {Lanzafame}, \& {Guinan}}]{Messina.2010}
{Messina}, S., {Desidera}, S., {Turatto}, M., {Lanzafame}, A.~C., \& {Guinan},
  E.~F. 2010, \aap, 520, A15

\bibitem[{{Metchev} {et~al.}(2005){Metchev}, {Eisner}, {Hillenbrand}, \&
  {Wolf}}]{Metchev.2005}
{Metchev}, S.~A., {Eisner}, J.~A., {Hillenbrand}, L.~A., \& {Wolf}, S. 2005,
  \apj, 622, 451

\bibitem[{{Meunier} {et~al.}(2012){Meunier}, {Lagrange}, \& {De
  Bondt}}]{Meunier.2012}
{Meunier}, N., {Lagrange}, A.-M., \& {De Bondt}, K. 2012, \aap, 545, A87

\bibitem[{{Mordasini} {et~al.}(2009){Mordasini}, {Alibert}, \&
  {Benz}}]{Mordasini.2009}
{Mordasini}, C., {Alibert}, Y., \& {Benz}, W. 2009, \aap, 501, 1139

\bibitem[{{Perryman} {et~al.}(2014){Perryman}, {Hartman}, {Bakos}, \&
  {Lindegren}}]{Perryman.2014}
{Perryman}, M., {Hartman}, J., {Bakos}, G.~{\'A}., \& {Lindegren}, L. 2014,
  \apj, 797, 14

\bibitem[{{Pontoppidan} {et~al.}(2008){Pontoppidan}, {Blake}, {van Dishoeck},
  {Smette}, {Ireland}, \& {Brown}}]{Pontoppidan.2008}
{Pontoppidan}, K.~M., {Blake}, G.~A., {van Dishoeck}, E.~F., {et~al.} 2008,
  \apj, 684, 1323

\bibitem[{{Press} \& {Rybicki}(1989)}]{Press.1989}
{Press}, W.~H. \& {Rybicki}, G.~B. 1989, \apj, 338, 277

\bibitem[{{Rameau} {et~al.}(2013){Rameau}, {Chauvin}, {Lagrange}, {Klahr},
  {Bonnefoy}, {Mordasini}, {Bonavita}, {Desidera}, {Dumas}, \&
  {Girard}}]{Rameau.2013}
{Rameau}, J., {Chauvin}, G., {Lagrange}, A.-M., {et~al.} 2013, \aap, 553, A60

\bibitem[{{Roberge} {et~al.}(2005){Roberge}, {Weinberger}, {Redfield}, \&
  {Feldman}}]{Roberge.2005}
{Roberge}, A., {Weinberger}, A.~J., {Redfield}, S., \& {Feldman}, P.~D. 2005,
  \apjl, 626, L105

\bibitem[{{Scargle}(1982)}]{Scargle.1982}
{Scargle}, J.~D. 1982, \apj, 263, 835

\bibitem[{{Setiawan} {et~al.}(2008){Setiawan}, {Henning}, {Launhardt},
  {M{\"u}ller}, {Weise}, \& {K{\"u}rster}}]{Setiawan.2008}
{Setiawan}, J., {Henning}, T., {Launhardt}, R., {et~al.} 2008, \nat, 451, 38

\bibitem[{{Sozzetti} {et~al.}(2014){Sozzetti}, {Giacobbe}, {Lattanzi},
  {Micela}, {Morbidelli}, \& {Tinetti}}]{Sozzetti.2014}
{Sozzetti}, A., {Giacobbe}, P., {Lattanzi}, M.~G., {et~al.} 2014, \mnras, 437,
  497

\bibitem[{{Suzuki} {et~al.}(2010){Suzuki}, {Kudo}, {Hashimoto}, {Carson},
  {Egner}, {Goto}, {Hattori}, {Hayano}, {Hodapp}, {Ito}, {Iye}, {Jacobson},
  {Kandori}, {Kusakabe}, {Kuzuhara}, {Matsuo}, {Mcelwain}, {Morino}, {Oya},
  {Saito}, {Shelton}, {Stahlberger}, {Suto}, {Takami}, {Thalmann}, {Watanabe},
  {Yamada}, \& {Tamura}}]{Suzuki.2010}
{Suzuki}, R., {Kudo}, T., {Hashimoto}, J., {et~al.} 2010, in \procspie, Vol.
  7735, Ground-based and Airborne Instrumentation for Astronomy III, 773530

\bibitem[{{Torres} {et~al.}(2006){Torres}, {Quast}, {da Silva}, {de La Reza},
  {Melo}, \& {Sterzik}}]{Torres.2006}
{Torres}, C.~A.~O., {Quast}, G.~R., {da Silva}, L., {et~al.} 2006, \aap, 460,
  695

\bibitem[{{van der Plas} {et~al.}(2015){van der Plas}, {van den Ancker},
  {Waters}, \& {Dominik}}]{vanderPlas.2015}
{van der Plas}, G., {van den Ancker}, M.~E., {Waters}, L.~B.~F.~M., \&
  {Dominik}, C. 2015, \aap, 574, A75

\bibitem[{{van Leeuwen}(2007)}]{Vanleeuwen.2007}
{van Leeuwen}, F. 2007, \aap, 474, 653

\bibitem[{{Vigan} {et~al.}(2012){Vigan}, {Patience}, {Marois}, {Bonavita}, {De
  Rosa}, {Macintosh}, {Song}, {Doyon}, {Zuckerman}, {Lafreni{\`e}re}, \&
  {Barman}}]{Vigan.2012}
{Vigan}, A., {Patience}, J., {Marois}, C., {et~al.} 2012, \aap, 544, A9

\end{thebibliography}

\begin{appendix} 
\section{}
We demonstrate here the formula of proportionality that connects the values of periodograms computed using two different planet masses, at the same semi-major axis. Let's consider two planets respectively with masses $m_{\rm 1}$ and $m_{\rm 2}$, orbiting the same star of mass $m_{\rm *}$ at the same period ${\cal P}$ and having the same orbital inclination $i$. The half-amplitudes of their RV curves are respectively:
\begin{eqnarray}
K_1=(\frac{2\pi G}{{\cal P}})^{\frac{2}{3}}\frac{m_{\rm 1}\sin i}{(m_{\rm *}+m_{\rm 1})^{\frac{2}{3}}}(1-e^2)^{-\frac{1}{2}}\\
K_2=(\frac{2\pi G}{{\cal P}})^{\frac{2}{3}}\frac{m_{\rm 2}\sin i}{(m_{\rm *}+m_{\rm 2})^{\frac{2}{3}}}(1-e^2)^{-\frac{1}{2}}
\end{eqnarray}
\\
Then, the ratio of these half-amplitudes is:
\begin{eqnarray}
\frac{K_1}{K_2}=\frac{m_{\rm 1}}{m_{\rm 2}}(\frac{m_{\rm *}+m_{\rm 2}}{m_{\rm *}+m_{\rm 1}})^{\frac{2}{3}}
\end{eqnarray}
\\
The Lomb-Scargle periodograms give the spectral power of the RV measurements as a function of the planet periods. Mathematically, the Lomb-Scargle periodograms are the square of the Fourier Transform of the RV measurements, we have:
\begin{eqnarray}
{\cal P_{\rm 1}}(P)\propto K_1^2 \\
{\cal P_{\rm 2}}(P)\propto K_2^2
\end{eqnarray}
Then, the ratio between the values of two periodograms for the period $P$ can be written:
\begin{eqnarray}
\frac{{\cal P_{\rm 1}}(P)}{{\cal P_{\rm 2}}(P)}=(\frac{m_{\rm 1}}{m_{\rm 2}}\times(\frac{m_{\rm *}+m_2}{m_{\rm *}+m_{\rm 1}})^{\frac{2}{3}})^2
\end{eqnarray}

\end{appendix}

\end{document}